\documentclass[
aps,
prd,
showpacs,
superscriptaddress,
nofootinbib,
floatfix]
{revtex4}
\usepackage{graphicx,
longtable}
\begin{document}
\title{Getting the most from the statistical analysis
of solar neutrino oscillations}
\author{        G.L.~Fogli}
\affiliation{   Dipartimento di Fisica
                and Sezione INFN di Bari\\
                Via Amendola 173, 70126 Bari, Italy\\}
\author{        E.~Lisi}
\affiliation{   Dipartimento di Fisica
                and Sezione INFN di Bari\\
                Via Amendola 173, 70126 Bari, Italy\\}
\author{        A.~Marrone}
\affiliation{   Dipartimento di Fisica
                and Sezione INFN di Bari\\
                Via Amendola 173, 70126 Bari, Italy\\}
\author{        D.~Montanino}
\affiliation{   Dipartimento di Scienza dei Materiali
                and Sezione INFN di Lecce\\
                Via Arnesano, 73100 Lecce, Italy\\}
\author{        A.~Palazzo}
\affiliation{   Dipartimento di Fisica
                and Sezione INFN di Bari\\
                Via Amendola 173, 70126 Bari, Italy\\}
\begin{abstract}
We present a thorough analysis of the current solar neutrino data,
in the context of two-flavor active neutrino oscillations. We aim
at performing an accurate and exhaustive statistical treatment of
both the input and the output information. Concerning the input
information, we analyze 81~observables, including the total event
rate from the chlorine experiment, the total gallium event rate
and its winter-summer  difference, the 44 bins of the
Super-Kamiokande (SK) energy-nadir electron spectrum, and the 34
day-night energy spectrum bins from the Sudbury Neutrino
Observatory (SNO) experiment. We carefully evaluate and propagate
the effects of 31 correlated systematic uncertainties, including
12 standard solar model (SSM) input errors, the $^8$B neutrino
energy spectrum uncertainty, as well as 11 and 7 systematics in SK
and SNO, respectively. Concerning the output information, we
express the $\chi^2$ analysis results in terms of ``pulls,''
embedding the single contributions to the total $\chi^2$ coming
from both the observables and the systematics. It is shown that
the pull method, as compared to the (numerically equivalent)
covariance matrix approach, is not only simpler and more
advantageous, but also includes useful indications about the
preferred variations of the neutrino fluxes with respect to their
SSM predictions. Our final results confirm the current best-fit
solution at large mixing angle (LMA), but also allow, with
acceptable statistical significance, other solutions in the
low-mass (LOW) or in the quasi-vacuum oscillation (QVO) regime.
Concerning the LMA solution, our analysis provides conservative
bounds on the oscillation parameters, and shows that the
contribution of correlated systematics to the total $\chi^2$ is
rather modest. In addition, within the LMA solution, the allowed
variations from SSM neutrino fluxes are presented in detail.
Concerning the LOW and QVO solutions, the analysis of the pull
distributions clearly shows that they are still statistically
acceptable, while the small mixing angle (SMA) solution could be
recovered only by {\em ad hoc\/} ``recalibrations'' of several SSM
and experimental systematics. A series of appendices elucidate
various topics related to the $\chi^2$ statistics, the
winter-summer difference in GALLEX/GNO, the treatment of the SK
and SNO spectra, and a quasi-model-independent comparison of the
SK and SNO total rates.
\end{abstract}
\medskip
 \maketitle

\section{Introduction}

The data from the Homestake \cite{Cl98}, SAGE \cite{Ab02},
GALLEX/GNO \cite{Ha99,Ki02}, Kamiokande \cite{Fu96},
Super-Kamiokande (SK) \cite{Fu02}, and Sudbury Neutrino
Observatory (SNO) \cite{AhCC,AhNC,AhDN} experiments have
consistently established that electron neutrinos emitted from the
Sun \cite{Ba89} undergo flavor transitions to the other active
states ($\nu_\mu$ or $\nu_\tau$). Neutrino oscillations
\cite{NuOs}, possibly affected by matter effects in the Sun or in
the Earth \cite{Matt}, represent a beautiful explanation of such
transitions.

Assuming the simplest scenario of two-family oscillations among
active neutrinos, an important task for the next future is to
refine the current constraints on the neutrino squared mass
difference $\delta m^2=m^2_2-m^2_1>0$ and on the mixing angle
$\theta_{12}\in[0,\pi/2]$. In order to accomplish this task, one
needs: (a) new or more precise measurements; (b) accurate
calculations of the $\nu_e$ survival probability $P_{ee}(\delta
m^2,\theta_{12})$ and of related observable quantities; and (c)
powerful statistical analyses to compare the (increasingly large)
solar $\nu$ data set with theoretical expectations.

The point (a), not discussed in this work, will soon be addressed
by the decisive reactor $\overline\nu_e$ experiment KamLAND
\cite{KamL}, as well as by the solar $\nu$ experiments which are
currently running \cite{Ab02,Ki02,AhDN}, being restored
\cite{Fu02}, or in construction \cite{BORE}. Concerning the point
(b), since the current numerical and analytical understanding of
the oscillation probability (and related observables) is quite
mature in the whole $(\delta m^2,\theta_{12})$ plane, we will only
make a few remarks when needed. In this paper, we rather focus on
point (c), aiming to an exhaustive statistical analysis including
all known observables and uncertainties in input, and providing
very detailed information in output, in order to better appreciate
the current status of the solutions to the solar $\nu$ problem in
terms of active flavor oscillations. Although some details will be
specific of solar $\nu$ data, the analysis method that we discuss
is quite general, and can be easily extended to any kind of global
fit.

The structure of our paper is the following. In Section~II we
discuss the equivalence between the $\chi^2$ approaches in terms
of the covariance matrix and of ``pulls'' of observables and
systematics. In Sec.~III we describe the input and output of the
pull approach, as applied to the analysis of 81 solar neutrino
observables and of 31 input systematics, in the context of $2\nu$
active oscillations. In Sec.~IV we discuss the $\chi^2$ analysis
results in terms of multiple allowed regions in the mass-mixing
parameter space $(\delta m^2,\tan^2\theta_{12})$ and in terms of
the associated pull distributions, which provide additional
information about the relative likelihood of the various solutions
and about the allowed deviations from standard solar model (SSM)
fluxes. We draw our conclusions in Sec.~V. More technical (but
sometimes substantial) issues are discussed in a series of
Appendices, which deal with the $\chi^2$ statistics
(App.~\ref{theorem}), the winter-summer asymmetry in GALLEX/GNO
(App.~\ref{wsapp}),  the treatment of the SK spectrum
uncertainties (App.~\ref{skspectrum}), the analysis of the SNO
data (App.~\ref{snotreat}), and a quasi-model-independent
comparison of SK and SNO total rates (App.~\ref{qmianalysis}).

As a conclusion to this Introduction, we would like to stress that
deepening the statistical analysis and improving the evaluation of
the uncertainties is an important task in neutrino oscillation
physics, just as it happens (or happened) in other areas of
``precision'' physics. Indeed, after the observation of two large
oscillation effects (the disappearance of atmospheric $\nu_\mu$
and of solar $\nu_e$, and their upcoming tests at long-baseline
accelerator and reactor experiments), we are likely to face an era
of delicate searches for smaller effects related, e.g., to the
angle $\theta_{13}$, to leptonic CP violation, or to subleading
contributions induced by nonstandard $\nu$ states or interactions.
Moreover, one should not forget that, so far, there is no {\em
direct\/} evidence for a vacuum oscillation pattern (disappearance
and reappearance of a specific flavour) or for matter effects in
the Sun or the Earth. Such effects might well generate only small
signals in present or planned experiments, and any effort should
be made in order to quantify them (if any) with accurate analyses.
From this viewpoint, we think that our thorough analysis can add
valuable information and useful technical tools to other solar
$\nu$ fits \cite{Ba02,AhDN,Cr01,Ch02,Pe02,Ho02,St02,Fu02} that
appeared soon after the release of the SNO neutral current data
\cite{AhNC}.

\section{Two equivalent ways of defining the $\chi^2$ function}

Let us consider a set of $N$ observables $\{R_{n}\}_{n=1,\dots,N}$
with their associated sets of experimental observations
$\{R_n^\mathrm{expt}\}$ and theoretical predictions
$\{R_n^\mathrm{theo}\}$. In general, one wants to build a $\chi^2$
function which measures the differences
$R_n^\mathrm{expt}-R_n^\mathrm{theor}$ in units of the total
(experimental and theoretical) uncertainties. This task is
completely determined if, for any difference
$R_n^\mathrm{expt}-R_n^\mathrm{theor}$, one can estimate an
uncorrelated error $u_n$, and a set of $K$ correlated systematic
errors $c^k_n$ induced by $K$ independent sources, namely
\begin{equation}\label{RexpRtheo}
R_n^\mathrm{expt}-R_n^\mathrm{theor}\pm u_n \pm c^1_n \pm c^2_n
\dots \pm c^K_n\ (n=1,\dots, N)\ ,
\end{equation}
with
\begin{eqnarray}
\rho(u_n,u_m) &=& \delta_{nm}\ , \label{corru}\\
\rho(c^k_n,c^h_m) &=& \delta_{kh}\ \forall (n,m)\ ,\label{corrc}
\end{eqnarray}
where $\rho$ represents the  correlation index.%
\footnote{The error $c_n^k$ represents the shift of the $n$-th
observable induced by a $+1\sigma$ variation in the $k$-th
systematic error source. Linear propagation of errors is assumed,
namely, possible $\pm 1\sigma$ asymmetries and second-order
systematic effects $\propto c^k_n c^{h}_n$ are consistently
neglected in computing the uncertainties of the difference
$R_n^\mathrm{expt}-R_n^\mathrm{theor}$.}

Given the input numbers in Eq.~(\ref{RexpRtheo}), two  $\chi^2$
definitions have been basically used in global analyses (with some
variants or combinations). They will be referred to as the
``covariance'' approach (Sec.~\ref{Capproach}) and the ``pull''
approach (Sec.~\ref{Papproach}). Although seemingly different, the
two approaches are strictly equivalent (Sec.~\ref{CeqP}).  The
pull approach, however, proves to be much more advantageous, and
will be used throughout this paper.

\subsection{The covariance approach}\label{Capproach}

In the ``covariance approach,'' one builds the (covariance) matrix
of squared errors as
\begin{equation}\label{sigmanm}
\sigma^2_{nm}=\delta_{nm}u_n u_m + \sum_{k=1}^K c^k_n c^k_m\ ,
\end{equation}
then inverts it, and evaluates the quadratic form
\begin{equation}\label{chicovardef}
\chi^2_\mathrm{covar}=\sum_{n,m=1}^N
(R_n^\mathrm{expt}-R_n^\mathrm{theor})[\sigma^2_{nm}]^{-1}
(R_m^\mathrm{expt}-R_m^\mathrm{theor})\ .
\end{equation}

This approach, proposed in \cite{Fo95} for the data available at
that time, has been later used in the majority of solar $\nu$
analyses of total events rates, with some variants related to the
treatment of the $^8$B $\nu$ flux (free or SSM) and to the
separation of spectral and total rate information in the SK data.

\subsection{The pull approach}\label{Papproach}

The alternative ``pull approach'' embeds the effect of each
independent $k$-th source of systematics through a shift of the
difference $R_n^\mathrm{expt}-R_n^\mathrm{theor}$ by an amount
$-\xi_k c^n_k$, where $\xi_k$ is a univariate gaussian random
variable%
\footnote{The minus sign preceding the terms $\xi_k c^k_n$ is
conventional. It amounts to attribute all the shifts to the
theoretical estimate:  $R_n^\mathrm{theor} \to R_n^\mathrm{theor}
+ \sum_{k=1}^K\xi_k c^k_n$.}
\begin{equation}\label{Rpull}
(R_n^\mathrm{expt}-R_n^\mathrm{theor}) \to
(R_n^\mathrm{expt}-R_n^\mathrm{theor}) - \sum_{k=1}^K\xi_k c^k_n\
.
\end{equation}
The normalization condition for the $\xi_k$'s is  implemented
through quadratic penalties in the $\chi^2$, which is then
minimized with respect to all $\xi_k$'s,
\begin{equation}\label{chipulldef}
\chi^2_\mathrm{pull}=\min_{\{\xi_k\}}\left[\sum_{n=1}^N
\left(\frac{R_n^\mathrm{expt}-R_n^\mathrm{theor}-
\sum_{k=1}^K\xi_k c_n^k}{u_n}\right)^2+\sum_{k=1}^K
\xi_k^2\right]\ .
\end{equation}
Denoting as $\overline \xi_k$ (``pulls'' of the systematics) the
values of the $\xi_k$'s at the minimum, and defining the ``pulls''
$x_n$ of the observables as
\begin{equation}\label{xbarn}
\overline x_n =\frac{R_n^\mathrm{expt}-(R_n^\mathrm{theor}+
\sum_{k=1}^K\overline\xi_k\, c_n^k)}{u_n}\ ,
\end{equation}
the value of $\chi^2_\mathrm{pull}$ is then split into two
diagonalized pieces, embedding the contribution to the $\chi^2$
from the residuals of the observables and of the systematics,
\begin{eqnarray}\label{obssys}
\chi^2_\mathrm{pull}&=& \chi^2_\mathrm{obs}+\chi^2_\mathrm{sys}\\
\label{chipulldiag} &=&\sum_{n=1}^N \overline x^2_n + \sum_{k=1}^K
\overline\xi^2_k\ .
\end{eqnarray}

The pull approach has often been used by the SK collaboration in
their day-night spectral analysis, in combination with the
covariance method for non-SK data \cite{Fu01}. More recently, the
SK energy-nadir spectrum has been analyzed through a mixed
pull-covariance approach \cite{Sm01}. The link between the
covariance and pull method, discussed in the next section, is also
mentioned in passing in Ref.~\cite{St01}. To our knowledge,
however, a complete analysis of solar neutrino data in terms of
$\chi^2_\mathrm{pull}$ has not been performed, prior to the
present work.

\subsection{Comparison and equivalence of the covariance and pull approaches}
\label{CeqP}

It is perhaps not generally known that, although seemingly
different, the covariance and pull approaches are strictly
equivalent,
\begin{equation}\label{chiequiv}
\chi^2_\mathrm{covar}\equiv \chi^2_\mathrm{pull}\ ,
\end{equation}
Our proof of the above identity is given in
Appendix~\ref{theorem}.%
\footnote{We have recently realized that Eq.~(\ref{chiequiv}) and
its implications have been also discussed \cite{Stum} and are
routinely used \cite{Bo02,Pu02} in the context of parton density
distribution fitting \cite{Th02}. Closely related results have
also been recently found in the context of cosmic microwave
background data fitting \cite{Br01}. Although the connection
between covariance matrix and pulls appears thus to be an
ubiquitous result in physics data analysis, we have been unable to
trace explicit references to Eq.~(\ref{chiequiv}) prior to
Ref.~\cite{Stum}.}
Given the equivalence in Eq.~(\ref{chiequiv}), the choice between
the covariance and the pull approach must be dictated by their
relative merits.

In particle physics, the covariance approach is typically used
either when the experimental collaborations provide detailed
information about the correlation matrix (as in the case of the
LEP Electroweak Working Group \cite{EWWG}), or when $N\ll K$ (as
in the case of solar neutrino fits to total rates only
\cite{Fo95}). However, for increasing $N$ the approach becomes
increasingly complicated. The inversion of large $N\times N$
covariance matrices, besides being numerically tricky, can make it
difficult to fully understand the results of global analyses.
Indeed, the current solar or atmospheric neutrino data fits,
involving $N\sim O(10^2)$, are getting close to their
manageability limits in terms of covariance matrices. The
situation might become even more problematic in future
high-statistics experiments, such as the neutrino factories from
muon storage rings or superbeams, where the oscillation parameters
will be inferred from the analysis and comparison of densely
binned and correlated (anti)neutrino event
spectra.%
\footnote{The pull approach has been recently applied to
prospective studies in this context \cite{Hu02}.}

The pull approach is clearly more practical than the covariance
one when $K\ll N$. In fact, the minimization in
Eq.~(\ref{chipulldef}) leads to a set of $K$ linear equations in
the $\overline \xi_k$'s, and to an associated $K\times K$ matrix
inversion, rather than the $N\times N$ covariance matrix inversion
(see App.~\ref{theorem}). Moreover, the final decomposition in
terms of pulls of observables and systematics [Eqs.~(\ref{obssys})
and (\ref{chipulldiag})] allows to trace the {\em individual\/}
contributions to the $\chi^2$, and
to easily detect anomalously large residuals.%
\footnote{The usefulness of the pull distribution diagnostics has
been already recognized in many areas of physics, including solar
neutrino physics \cite{Kr02,Ho02} and electroweak precision
physics \cite{EWWG}. However, in Refs.~\cite{Kr02,Ho02,EWWG} the
pulls are defined without the shifts in Eq.~(\ref{xbarn}), and are
thus all correlated to each other.}
In general, this method is useful to gauge the mutual agreement of
data in a global fit, or to diagnose tension among data (if any),
for any given point in the model parameter space [($\delta
m^2,\tan^2\theta_{12}$) in our case]. The analysis of the
$\chi^2_\mathrm{obs}$ and $\chi^2_\mathrm{sys}$ components
[Eq.~(\ref{obssys})] can also be useful to trace possible sources
of good or bad fits.

Given the advantages of $\chi^2_\mathrm{pull}$ in cases where
$K\ll N$, we have redesigned the statistical analysis of solar
neutrino oscillations in terms of pulls, and applied it to the
current data set (where $N=81$ and $K=31$), as discussed in the
next section.

\section{Input and output for the $\chi^2_\mathrm{pull}$ analysis}

In this section we describe the main input and output quantities,
related to the $\chi^2_\mathrm{pull}$ analysis of solar neutrino
data. In input we consider a set of $N=81$ observables $R_n$ (with
associated uncorrelated errors $u_n$), and a set of $K=31$ sources
of correlated systematic errors $c^k_n$, in part related to the
SSM and in part to the experiments.%
\footnote{Systematic error sources are defined as ``correlated''
if they act upon two or more observables at the same time.
Systematics which act upon one observable only (e.g., the $\nu$-Cl
absorption cross section error) simply contribute quadratically to
the uncorrelated error for that observable (the Cl total rate, in
the example).}
In output we consider the total $\chi^2_\mathrm{pull}$, its
decomposition in individual pulls, and the shifts of the neutrino
fluxes from their SSM value.

\subsection{Input observables and uncorrelated errors}
\label{InputO}

The first two observables in our list are the Chlorine total rate
\cite{Cl98},
\begin{equation}\label{Clrate}
R_\mathrm{Cl}^\mathrm{expt}=2.56 \pm 0.23 \mathrm{\ SNU}\ ,
\end{equation}
and the average Gallium total rate (SAGE \cite{Ab02} + GALLEX/GNO
\cite{Ki02}),
\begin{equation}\label{Garate}
  R_\mathrm{Ga}^\mathrm{expt}=70.8\pm 4.4 \mathrm{\ SNU}\ ,
\end{equation}
where the errors include the statistical and experimental
systematic contributions to the uncorrelated errors
$u_\mathrm{Cl}$ and $u_\mathrm{Ga}$, respectively. In the
analysis, the corresponding cross section uncertainties must also
added in quadrature. Following the suggestion in \cite{Go02}, the
cross section error components $\Delta R_{X,i}$ for $X=$(Cl, Ga)
are first added linearly and then quadratically into low ($L$) and
``high'' ($H$) energy parts [$L=$(pp, pep, Be, N, O) and $H=$(B,
hep), respectively]%
\footnote{The prescription in Eq.~(\ref{cserrors}) is intermediate
between the extreme possibilities of quadratic sum \cite{Fo95} and
linear sum \cite{Ga00} over all flux components. We observe that
this prescription is not only justified by the physics of the
$\nu_e$ absorption processes in Cl and Ga \cite{Go02}, but also by
the effective separation of the Cl and the Ga solar $\nu$ response
functions into two ``$L$'' and ``$H$'' clusters in the energy
domain \cite{Ba91,Bh98}.}
\begin{equation}
\label{cserrors}
 u^2_X(\mathrm{cross\ section})=
 \left(\sum_{i\in L} \Delta R^\mathrm{theo}_{X,i}\right)^2 +
\left(\sum_{i\in H} \Delta R^\mathrm{theo}_{X,i} \right)^2\ .
\end{equation}
For $X$=Cl, the cross section error components are evaluated as
$\Delta R_{X,i}=R_{X,i}\,\Delta \ln C_{X,i}$, where $R_{X,i}$ are
the (oscillated) rate components, and the fractional $1\sigma$
cross section uncertainties $\Delta\ln C_{X,i}$  can be taken from
the compilation in \cite{Li00}. For $X$=Ga, the value of $\Delta
R_{X,i}$ is computed by taking 1/3 of the variations induced by
the $\pm 3\sigma$ perturbed cross sections \cite{Ba97} on the
$i$-th Ga rate component for each point of the oscillation
parameter space, as suggested in \cite{Go02}.%
\footnote{We conservatively assume the largest between the
$+1\sigma$ and $-1\sigma$ asymmetric Ga cross section errors.}

Our third observable is the winter-summer ($W-S$) rate difference
\cite{Fo00} measured in GALLEX/GNO \cite{Ki02}, here introduced
for the first time in the oscillation analysis. This datum is
described in detail in App.~\ref{wsapp}.

The SK experiments provides 44 observables, in terms of (binned)
absolute event rates for the energy-nadir differential spectrum of
electrons \cite{Fu02}. Our treatment of the SK spectral
information is described in detail in Sec.~\ref{skspectrum}.

The set of solar $\nu$ observables is completed by the 34
day-night energy spectrum bins from the SNO experiment
\cite{AhNC,AhDN}, which include contributions from $\nu$ elastic
scattering (ES), charged current (CC) and neutral current (NC)
interactions, and backgrounds \cite{HOWT}. Our treatment of the
SNO spectral information is described in App.~\ref{snotreat}.

\subsection{Input correlated systematics}
\label{InputS}

The 31 sources of correlated systematics include 12 uncertainties
related to SSM input, the $^8$B $\nu$ energy shape uncertainty, 11
SK error sources and 7 SNO error sources.

Concerning the SSM input, we take from \cite{Ba01} the central
values for the fluxes ($\Phi_\mathrm{pp}$, $\Phi_\mathrm{pep}$,
$\Phi_\mathrm{Be}$, $\Phi_\mathrm{B}$, $\Phi_\mathrm{N}$,
$\Phi_\mathrm{O}$), but rescale $\Phi_\mathrm{hep}$ from the value
in \cite{Ba01} ($9.3\times 10^3$ cm$^{-2}$~s$^{-1}$) to our
default value
\begin{equation}
\label{hepflux} \Phi_\mathrm{hep}=8.3 \times 10^3 \mathrm{\
cm}^{-2}\mathrm{\ s}^{-1}\ ,
\end{equation}
according to the recent evaluation of the associated
$S_\mathrm{hep}$ factor in \cite{Pa01}.

The SSM also embeds a set of eleven sources $X_k$ of correlated
uncertainties (the cross section factors $S_{11}$, $S_{13}$,
$S_{34}$, $S_{1,14}$, $S_{17}$, the Be capture cross section
$C_{\rm Be}$,  the Sun luminosity, metallicity $Z/X$, age,
opacity, and element diffusion), with fractional uncertainties
$\Delta \ln X_k$, as listed in \cite{Li00}. With respect to the
compilation in \cite{Li00}, we update $\Delta\ln Z/X=0.061$ from
\cite{Ba01}, and we add $X_{12}=S_\mathrm{hep}$,
with $\Delta \ln S_\mathrm{hep}=0.3$.%
\footnote{The authors of \cite{Pa01} quote an uncertainty of $\sim
15\%$ for $S_\mathrm{hep}$, that we conservatively double to
$30\%$.}
The effects of such sources of uncertainties on the neutrino
fluxes $\Phi_i$ are characterized by log-derivatives
\cite{Ba89,Fo95}, $\alpha_{ik}=\partial\ln\Phi_i/\partial\ln X_k
$, as compiled in \cite{Li00}. Concerning $X_{12}$, the only
nonzero log-derivative is $\alpha_{\mathrm hep ,12}=1$.

The collective effect of the  SSM sources of systematics $X_k$
amounts to shift the neutrino fluxes as
\begin{equation}\label{fluxshift}
 \Phi_i \to \Phi_i \left(1 + \sum_{k=1}^{12} \xi_k\, \alpha_{ik}\, \Delta\ln
 X_k\right)\ ,
\end{equation}
where the $\xi^k$'s, penalized by the quadratic term
$\sum_{k=1}^{12}\xi_k^2$ in the expression of
$\chi^2_\mathrm{pull}$, are  minimized away in the fit. Notice
that the above equation is the linearized form of the power laws
connecting each flux to the $X_k$'s \cite{Ba89,Ca97}. Such linear
form satisfies the luminosity constraint for the fluxes
\cite{Lumi} by construction, due to the sum rule discussed in
Ref.~\cite{Fo95,Li00} (see also \cite{Ul96}).

Within the pull approach, the shifts in Eq.~(\ref{fluxshift}) are
easily propagated to all theoretical predictions
$R_n^\mathrm{theo}$. In particular, if $R_{n,i}^\mathrm{theo}$ is
the $i$-th flux contribution to $R_n^\mathrm{theo}$, then the
associated correlated shift from the $k$-th source is $\xi_k
c^k_{n,i}=\xi_k R_{n,i}^\mathrm{theo}\alpha_{ik}\Delta\ln X_k$.
The net effect of the shifts in Eq.~(\ref{fluxshift}) is thus the
generation of correlated errors on the $R_n$'s, which is strictly
equivalent to the construction of the astrophysical error matrix
defined in the earlier covariance approach \cite{Fo95,Li00}, as
also noticed in \cite{St01}.

The 13th source of correlated systematics in our list is the $^8$B
$\nu$ spectrum shape uncertainty \cite{Boro} around the current
``central'' spectrum \cite{Or00}, which affects all the 81
observables%
\footnote{With the possible exception of the Ga winter-summer
difference, where its effects cancel to a large extent, and can be
safely neglected as compared with the rather large statistical
error (see App.~\ref{wsapp}).}
at the same time. In the absence of oscillations, we estimate that
a $+1\sigma$ perturbation of the $^8$B $\nu$ spectrum (in the
direction of higher $\nu$ energies) generates a $+2.2\%$ and a
$+1.7\%$ increase of the $\Phi_B$ component of the $R_\mathrm{Ga}$
and $R_\mathrm{Cl}$ theoretical rates, respectively (the
evaluation is repeated in each point of the mass-mixing plane).
The corresponding shifts for the SK and SNO spectra are evaluated
in App.~\ref{skspectrum} and \ref{snotreat}, respectively.

There are also eleven sources of correlated systematics, which
affect only the SK spectrum. They include the SK energy scale and
resolution uncertainties, an overall SK rate offset, and eight
sources of systematics separately affecting the eight energy bins,
with full correlation in nadir \cite{Fu02} (see
App.~\ref{skspectrum}).

Finally, there are seven sources of correlated systematics, which
affect only the SNO spectrum. They include: the uncertainties
affecting the SNO energy scale and resolution, the event vertex
reconstruction, the neutron capture efficiency, the neutron and
low-energy (LE) background estimates and the interaction cross
sections. See App.~\ref{snotreat} for more details.

\subsection{Output}

As output of the pull analysis, we get the function
$\chi^2_\mathrm{pull}(\delta m^2,\tan^2\theta_{12})$ (essential to
identify absolute and local minima and to draw confidence level
contours), as well as other useful statistical indicators.

Concerning $\chi^2_\mathrm{pull}$, for any {\em fixed\/} point in
the parameter space $(\delta m^2,\tan^2\theta_{12})$, the
goodness-of-fit test requires $\chi^2_\mathrm{pull}\sim N$
($N=81$) for an acceptable fit. Further information can be gained
by splitting $\chi^2_\mathrm{pull}$ into the separate
contributions $\chi^2_\mathrm{obs}$ and $\chi^2_\mathrm{sys}$
[Eqs.~\ref{obssys} and \ref{chipulldiag}], obtained by summing up
the squared pulls of the $N=81$ observables [$\overline x_n$, see
Eq.~(\ref{xbarn})] and of the $K=31$ systematics
[$\overline\xi_k$, see Eq.~(\ref{csibar})]. The larger the value
of $\chi^2_\mathrm{sys}$, the more the fit tends to ``stretch''
one or more correlated systematics to get a better agreement
between data and expectations. Apart from global features, the
analysis of the pull sets $\{\overline x_n\}$ and
$\{\overline\xi_k\}$  allows to quantify individual contributions
to the $\chi^2$, which, if anomalously large, might be indicative
of problems either in the theoretical predictions or in the
experimental measurements. Therefore, we think it useful to
present, besides the global values of $\chi^2_\mathrm{pull}=
\chi^2_\mathrm{obs}+\chi^2_\mathrm{sys}$, also some selected lists
of pulls.

Finally, it is useful to isolate the twelve SSM systematic pulls
$\{\overline\xi_k\}_{k=1,\dots,12}$ which, on the basis of
Eq.~(\ref{fluxshift}), allow to derive the induced neutrino flux
shifts from the SSM central values, namely
\begin{equation}\label{shiftflux}
 \frac{\Delta \Phi_i}{\Phi_i}=\sum_{k=1}^{12}\overline \xi_k
 \alpha_{ik}\Delta\ln X_k\ .
\end{equation}
These shifts provide valuable (and luminosity-constrained)
information about the preferred departures from the SSM  within
the various oscillation solutions to the solar neutrino problem.

Summarizing, we will show and discuss results about pulls,
\begin{eqnarray}
\label{firsteq}
\{\overline x_n\}_{n=1,\dots,81} &=& \mathrm{pulls\ of\ the\ observables}\ ,\\
\{\overline \xi_k\}_{k=1,\dots,31} &=& \mathrm{pulls\ of\ the\
correlated\ systematics}\ ,
\end{eqnarray}
about $\chi^2$ values,
\begin{eqnarray}
\chi^2_\mathrm{obs} &=& \sum_{n=1}^{81} \overline x_n^2\ ,\\
\chi^2_\mathrm{sys} &=& \sum_{k=1}^{31} \overline \xi_k^2\ ,\\
\chi^2_\mathrm{pull} &=& \chi^2_\mathrm{obs}+ \chi^2_\mathrm{sys}
\ , \label{lasteq}
\end{eqnarray}
and about fractional shifts from the SSM predictions,
\begin{equation}
\Delta\Phi_i/\Phi_i = \nu\mathrm{\ flux\ shifts}\ .
\end{equation}

\section{Results of the $\chi^2_\mathrm{pull}$ analysis}

In this section we start by describing the global results of the
$\chi^2_\mathrm{pull}$ analysis, and then we break down such
results at increasing levels of detail.

\begin{table}[t]
\caption{\label{chisquares} Positions and values of the absolute
minimum (LMA) and of three relevant local minima (LOW, QVO, SMA)
of $\chi^2_\mathrm{pull}$, together with the separate
contributions from pulls of observables ($\chi^2_\mathrm{obs}$)
and of correlated systematics ($\chi^2_\mathrm{sys}$). The
corresponding $\Delta\chi^2$ variations are also given.}
\begin{ruledtabular}
\begin{tabular}{llcrrrr|rr}
Solution & $\delta m^2$ (eV$^2$)& $\tan^2\theta_{12}$ &
$\chi^2_\mathrm{obs}$ & $\Delta \chi^2_\mathrm{obs}$ &
$\chi^2_\mathrm{sys}$& $\Delta\chi^2_\mathrm{sys}$
& $\chi^2_\mathrm{pull}$& $\Delta\chi^2_\mathrm{pull}$ \\
\hline
LMA &  $5.5\times 10^{-5}$ & 0.42 &                 71.3 & --- & 2.1 & ---  &73.4 & ---\\
LOW &  $7.3\times 10^{-8}$ & 0.67 &                 79.7 & 8.4 & 4.1 & 2.0  &83.8 & 10.4\\
QVO &  $6.5\times 10^{-10}$& 1.33  &                74.9 & 3.6 & 6.3 & 4.2  &81.2 & 7.8\\
SMA &  $5.2\times 10^{-6}$ & $1.1\times 10^{-3}$ &  83.1 &11.8 & 13.8& 11.7 &96.9 & 23.5\\
\end{tabular}
\end{ruledtabular}
\end{table}

\subsection{Global results}

The global results of our solar $\nu$ oscillation fit are
summarized in Table~\ref{chisquares} and in Fig.~\ref{fig01}. In
Table~\ref{chisquares} we report the $(\delta
m^2,\tan^2\theta_{12})$  coordinates of the best-fit point
[so-called large mixing angle (LMA) solution] and of the three
deepest (local) $\chi^2_\mathrm{pull}$ minima in the regions of
low $\delta m^2$ (LOW), quasivacuum oscillations%
\footnote{We do not find acceptable solutions in the
octant-symmetric vacuum oscillation (VO) regime. For the QVO
solution in Table~\ref{chisquares}, only the highest energy $\nu$
flux components ($\Phi_B$ and $\Phi_\mathrm{hep}$) have reached
the VO regime, while the lowest energy ones are still affected by
octant-asymmetric quasivacuum effects \cite{QVOs,Petc} in the
Sun.}
(QVO), and small mixing angle (SMA).

Concerning the goodness-of-fit test, we remind that, at the
absolute minimum, one expects the total $\chi^2$ to be in the $\pm
1\sigma$ range $N_\mathrm{DF}\pm\sqrt{ 2N_\mathrm{DF}}$
\cite{Revi}. In our case ($N_\mathrm{DF}=81-2$), it is
$\chi^2_\mathrm{pull}=73.4$ at the LMA best-fit point, well within
the expected range $79\pm 12.6$. Also the LOW and especially the
QVO solutions have acceptable values of $\chi^2_\mathrm{pull}$,
while the SMA value appears to be significantly larger than
expected. The pull analysis will confirm that the LOW and QVO
solutions are still viable, while the SMA solution is no longer
statistically acceptable. Notice that at the LMA point, most of
the contribution to $\chi^2_\mathrm{pull}$ comes from pulls of
observables ($\chi^2_\mathrm{obs}$) rather than systematics
($\chi^2_\mathrm{sys}$). All the other solutions in
Table~\ref{chisquares} show an increase of both
$\chi^2_\mathrm{pull}$ and $\chi^2_\mathrm{sys}$, implying an
increasing departure of the theoretical predictions from the data
and of the systematics offsets from zero.

Concerning the parameter estimation test%
\footnote{Useful discussions of the applications and differences
between the goodness-of-fit test and the parameter estimation test
can be found in \cite{Revi,Good,Ly99}.}
(based on $\Delta\chi^2_\mathrm{pull}$ variations around the
minimum) Figure~1 shows the results of our global analysis in the
usual mass-mixing plane. The confidence level isolines are drawn
at $\Delta\chi^2=4.61$, 5.99, 9.21, and 11.83, corresponding to
90\%, 95\%, 99\% and 99.73\% joint probability regions for the two
$(\delta m^2,\tan^2\theta_{12})$ parameters. The QVO and LOW
parameters are still acceptable at the 99\% and 99.73\% C.L.,
respectively, while the SMA parameters are basically ruled out.
Our LMA bounds appear to be (sometimes significantly) more
conservative than in other recent analyses
\cite{AhDN,Ba02,Cr01,Ch02,Pe02,Ho02,St02,Fu02}.%
\footnote{The closest agreement is reached with the global allowed
LMA region in Ref.~\cite{Pe02}.}
In particular, at the 99.73\% C.L.\ we derive from
Fig.~\ref{fig01} that: (i) maximal mixing is marginally allowed in
the LMA region, and (ii) the highest $\delta m^2$ allowed values
hit the region independently
disfavored by CHOOZ data \cite{CHOO}.%
\footnote{CHOOZ data are not included in the present analysis, in
order to show more clearly the strength of the upper bound on
$\delta m^2$ placed by solar neutrino data alone.}
We think that the detailed treatment of all known uncertainties
(and of their propagation to all relevant experimental
observables) plays a role in such different results, also for
non-LMA solutions. Concerning the LOW solution, we note that the
inclusion of the winter-summer datum from GALLEX/GNO contributes
to decrease its likelihood in our analysis.

\subsection{Separating experimental bounds}

\begin{figure}
\includegraphics[scale=0.90, bb= 100 100 510 750]{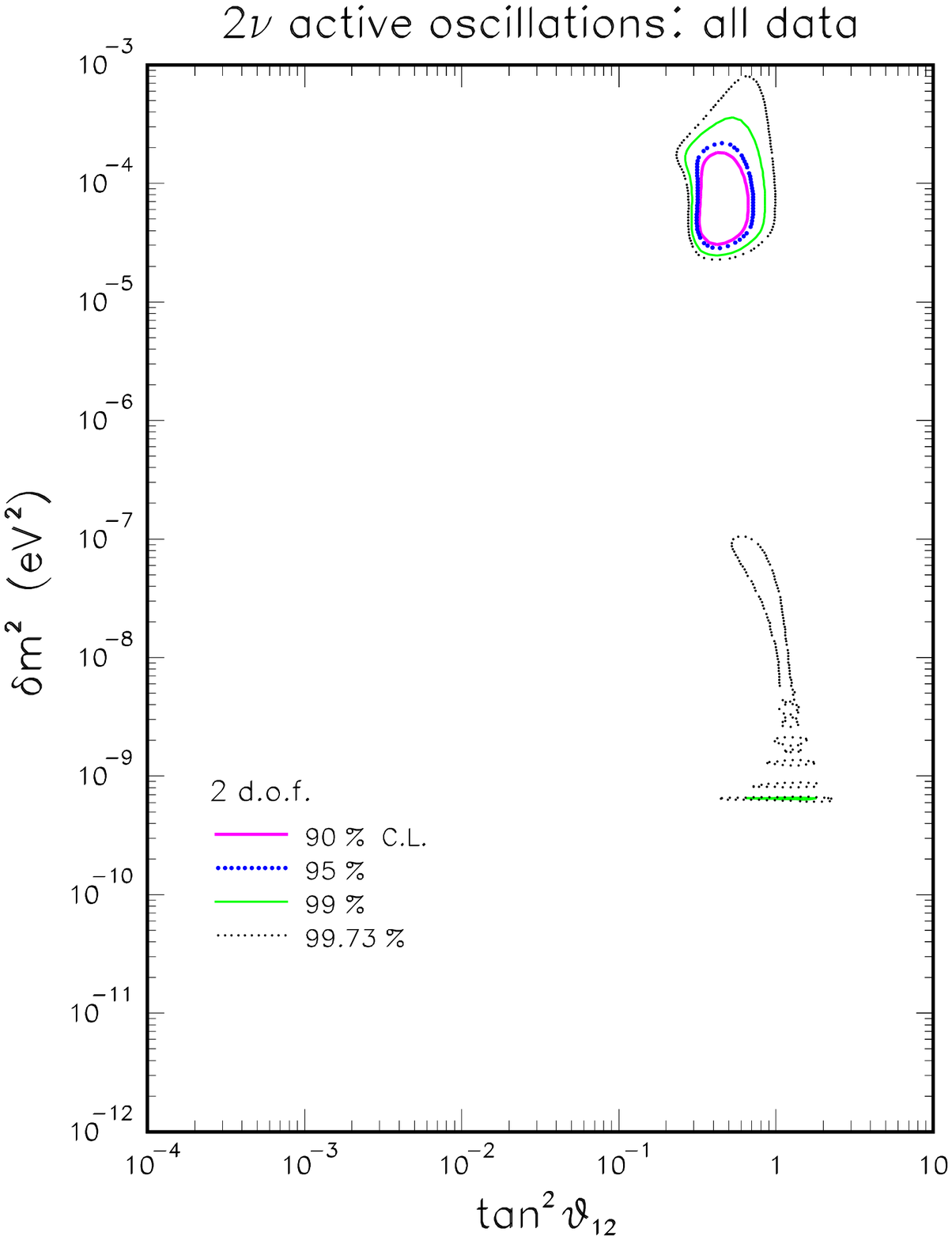}
\vspace*{+0cm} \caption{ \label{fig01} Global results of the solar
neutrino data analysis, including 81 observables and 31 sources of
correlated systematics. The parameter space $(\delta
m^2,\tan^2\theta_{12})$ refers to the scenario of $2\nu$
oscillations among active states. The relevant $\chi^2$ minima in
the LMA, LOW, and QVO regions are given in
Table~\ref{chisquares}.}
\end{figure}

\begin{figure}
\includegraphics[scale=0.90, bb= 100 100 510 750]{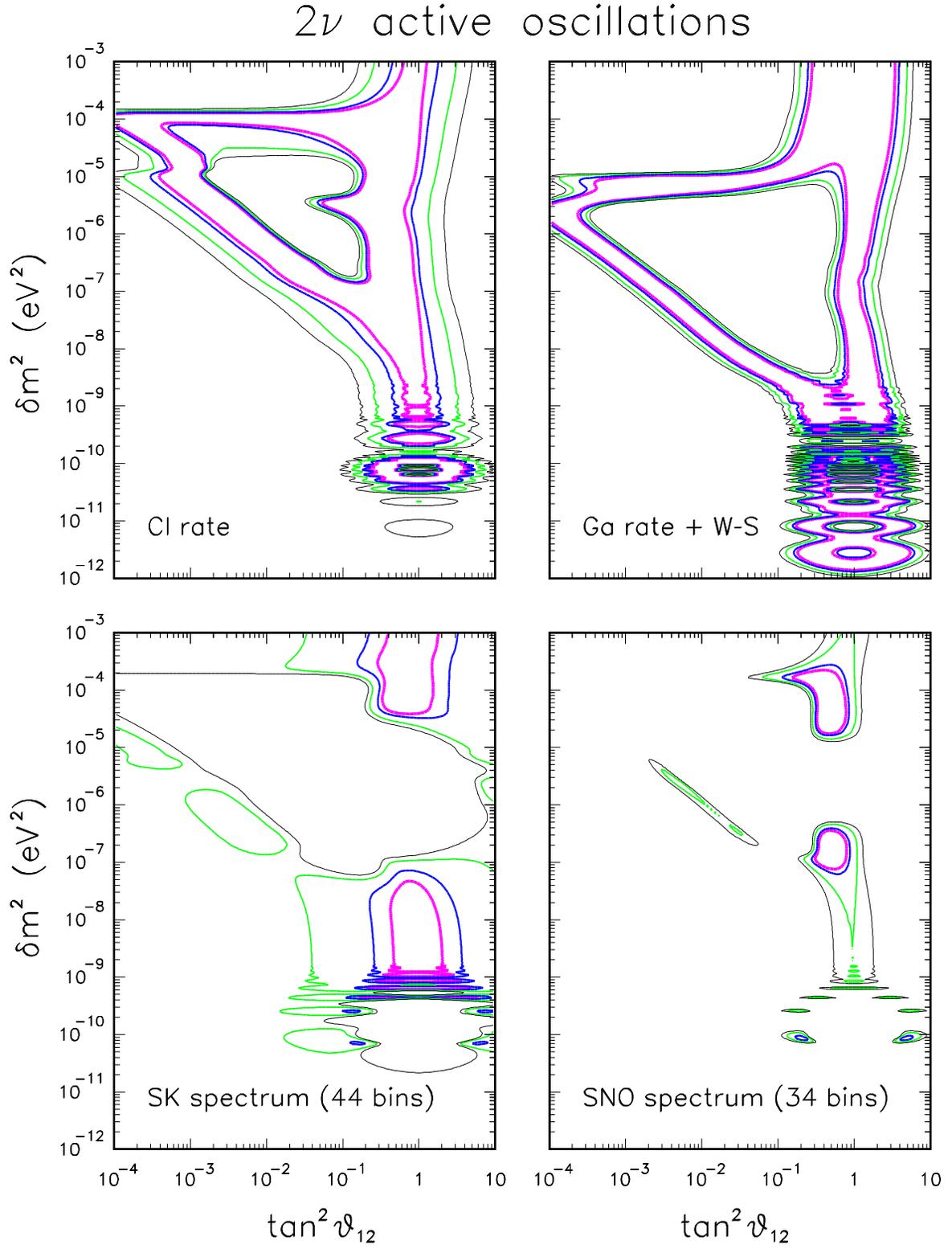}
\vspace*{+0.5cm} \caption{\label{fig02} Results of the solar
neutrino data analysis, as obtained by separating four classes of
observables: (i) the chlorine rate; (ii) the average
SAGE+GALLEX/GNO gallium rate plus the GALLEX/GNO winter-summer
difference; (iii) the SK energy-nadir spectrum; and (iv) the SNO
day-night spectrum.}
\end{figure}

Figure~\ref{fig02} show the decomposition of the global results
into contributions from the Cl experiment (total rate), from the
Gallium experiments (total rate and winter-summer difference),
from the SK energy-nadir spectrum (44 bins), and from the SNO
day-night spectrum (34 bins). In each panel, the results are shown
in terms of allowed regions, for the same confidence levels as in
Fig.~\ref{fig01} (referred to the absolute minimum in each panel).

Concerning the $\chi^2_\mathrm{pull}$ minima in Fig.~\ref{fig02},
their positions are not particularly interesting for the Cl and Ga
cases, where they are essentially degenerate. More interesting is
the case of the SK experiment alone, where the best fit
($\chi^2_\mathrm{pull}=38.4$) is  reached at maximal mixing and
for $\delta m^2=6.5\times 10^{-10}$ eV$^2$ (in agreement with the
results in \cite{Smy0}), very close to the QVO coordinates in
Table~\ref{chisquares}. Concerning the fit to SNO data only, we
find the best fit at $\delta m^2=3.7\times 10^{-5}$ eV$^2$ and
$\tan^2\theta_{12}=0.47$, close to the LMA coordinates in
Table~\ref{chisquares}, with $\chi^2_\mathrm{pull}=25.7$. The
latter value appears to be on the lower side of the $\pm1\sigma$
expected range for the $\chi^2_\mathrm{pull}$ in SNO ($32\pm\sqrt{64}$).%
\footnote{We think that this feature might be partly due to
non-optimal binning of the SNO spectrum. Although, at low energy,
relatively dense binning is required to enhance the effects of the
neutral current component, at high energies it is preferable to
enlarge the bin width so as to match the SNO energy resolution
width (analogously to the current SK energy spectrum binning).}

Concerning the shapes of the allowed regions in Fig.~\ref{fig02},
we note the following facts. None of the experiments excludes
maximal mixing and $\delta m^2\to\infty$ at 99\% C.L. This feature
is rather well known for Ga and SK data, but does not appear in
all recent analyses for the Cl and SNO data. For instance, the Cl
contours in Ref.~\cite{Fu02} appear to be more restrictive than
ours, which might be due to different estimates of the Cl errors.
We also note some differences between our SNO bounds in
Fig.~\ref{fig02} and the SNO official analysis in Fig.~4a of
\cite{AhDN}: (a) in the QVO region, our contours are smooth (as
they should); (b) we do not find Q(VO) solutions at maximal mixing
for $\delta m^2\sim 10^{-10}$ eV$^2$; (c) our bounds in the LMA
region allow maximal mixing and $\delta m^2\to\infty$ well within
the 99.73\% C.L. Concerning the points (a) and (b), we think that
the differences might depend in \cite{AhDN} on the (numerically
delicate) averaging of the oscillating terms in the energy or time
domain. Concerning the points (b) and (c), some differences might
also be due to the fact that the analysis in Fig.~4a of
\cite{AhDN} is done without SSM input. Concerning the point (c),
we note that the exclusion of $\delta m^2\to\infty$ {\em and\/} of
$\tan^2\theta_{12}=1$ at the 99.73\% level by the SNO data alone
(as found in Fig.~4a of \cite{AhDN}) would be equivalent to the
exclusion of the constant $P_{ee}=1/2$ case (or of equal $\nu_e$
and $\nu_{\mu,\tau}$ fluxes, $\Phi_e=\Phi_{\mu,\tau}$) at the same
confidence level. Although the SNO data clearly prefer $P_{ee}\sim
1/3$ \cite{AhNC} (see also App.~\ref{qmianalysis}), a $3\sigma$
rejection of $P_{ee}\simeq 1/2$ might be premature. Indeed, from
Fig.~3 of \cite{AhNC} it appears that the $\Phi_e=\Phi_{\mu,\tau}$
line touches the 95\% error ellipse determined by the total SNO
rates. Given that the exclusion of relatively large values of
$\delta m^2$ and $\tan^2\theta$ has profound implications in
lepton physics (both phenomenologically and theoretically), we
think that the impact of the SNO data on such values warrants
further investigations.

From Fig.~\ref{fig02} it also appears that all experiments largely
agree in the LMA region, and that a few  QVO ``islands'' below
$10^{-9}$ eV$^2$ also happen to be consistent with all
experiments. Moreover, all the experiments appear to be
generically consistent with some ``LOW'' or ``SMA'' regions at
least at 99\% C.L. However, such regions are somewhat different
for the different panels in Fig.~\ref{fig02}. Concerning the LOW
case, there is a reasonable overlap of the Cl, Ga, and SK bounds
at $\delta m^2\lesssim 10^{-7}$ eV$^2$, while SNO prefers $\delta
m^2\gtrsim 10^{-7}$ eV$^2$, where the LOW parameters more easily
adapt to the preferred value $P_{ee}\simeq 1/3$. This tension
generates an overall decrease of the LOW likelihood. Concerning
the SMA case, the low-$\tan^2\theta_{12}$ regions separately
allowed by Cl, Ga, SK, and SNO at the 99.73\% C.L.\ do not
overlap.

\begin{figure}
\includegraphics[scale=0.90, bb= 100 100 510 750]{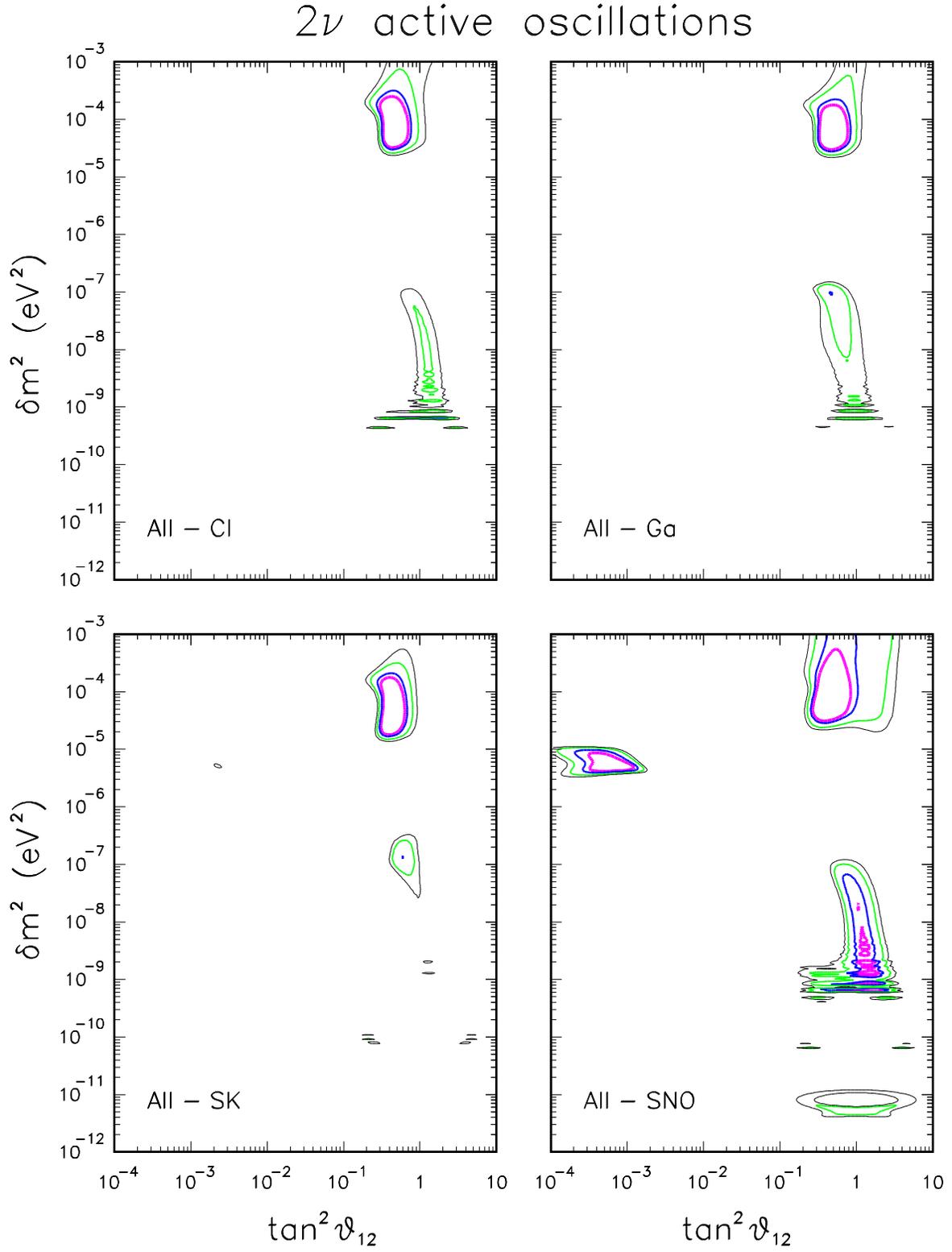}
\vspace*{+0.5cm} \caption{\label{fig03} Results of the solar
neutrino data analysis, as obtained by excluding each of the four
data sets in Fig.~\ref{fig02} from the global set used in
Fig.~\ref{fig01}.}
\end{figure}

Figure~\ref{fig03} allows to appreciate the impact of each
experiment in the global fit, by removing one experiment at a
time. In a sense, Fig.~\ref{fig03} is the ``difference'' between
Fig.~\ref{fig01} and Fig.~\ref{fig02}. It can be seen that the
removal of either the Cl or the Ga experiments (upper panels in
Fig.~\ref{fig03}) weakens the bounds on large mixing, on large
values of $\delta m^2$, and on the LOW parameters, but does not
alter the
situation for vacuum oscillations, which are excluded in both cases.%
\footnote{In Ref.~\cite{St02}, the Ga impact on the LOW solution
has been studied by lowering the total rate from $70.8\pm 4.4$ to
$66.1\pm 5.3$ SNU. The authors of \cite{St02} find a corresponding
reduction of the $\Delta\chi^2_\mathrm{LOW}$ from 6.9 to 3.0, with
respect to the LMA minimum. By repeating the same exercise, we
find a smaller effect: our $\Delta\chi^2_\mathrm{LOW}$ decreases
from 10.4 (see Table~\ref{chisquares}) to 9.0.}
The lower left panel in Fig.~\ref{fig03} shows that the removal of
the SK experiment would diminish  the likelihood of the solutions
characterized by no or mild energy spectrum distortion (e.g.,
$\delta m^2\to\infty$ or QVO cases), and, conversely, would make
some strongly energy-dependent solutions marginally reappear (at
small mixing and in the vacuum regime). Finally, the lower right
panel in Fig.~\ref{fig03} shows the pre-SNO situation (but with
updated SK, Cl, and Ga data), with all the well-known multiple
solutions to the solar neutrino problem.

The comparison of Figs.~\ref{fig01}--\ref{fig03} shows the
dramatic impact of SK and SNO in determining the preference for
large mixing (and especially for the LMA solutions) and the
rejection of the SMA solution. However, the Cl and Ga data still
play an important role in determining the shape of the LMA
contours, as well as the likelihood of the less favored solutions
(LOW and QVO), which cannot be rejected on the basis of the
present global information.

\subsection{Separating and grouping pulls}

We discuss the decomposition of $\chi^2_\mathrm{pull}$ into
separate and grouped pulls of observables $\{\overline{x}_n\}$ and
of correlated systematics $\{\overline{\xi}_k\}$ [see
Eqs.~(\ref{firsteq})--(\ref{lasteq})] for the various solutions.

\begingroup \squeezetable
\begin{table}[th]
\setlength\LTleft{0pt} \setlength\LTright{0pt}
\begin{longtable}{@{\extracolsep{30pt}}clrcrrr}
\caption{\label{obspulls} Pulls of the observables in the various
solutions.}\\[-2mm]
\hline\hline\\
n & Observable&LMA& LMA [min,~max] at $3\sigma$ & LOW & QVO & SMA\\ %
\hline
1 & Cl rate                           &$-1.74$&$[-2.56,\,-1.17]$&$-1.86$&$-2.68$&$ 0.14$\\ %
2 & Ga rate                           &$ 0.39$&$[-1.49,\,+1.90]$&$ 1.27$&$ 0.84$&$-0.19$\\ %
3 & Ga $W-S$                          &$-1.23$&$[-1.23,\,-1.22]$&$-1.75$&$-1.25$&$-1.22$\\ %
4 & SK [5.0,5.5]                      &$-0.09$&$[-0.46,\,+0.10]$&$ 0.14$&$-0.01$&$ 0.58$\\ %
5 & SK [5.5,6.5] day                  &$-0.36$&$[-1.06,\,+0.40]$&$-0.06$&$-0.52$&$ 0.17$\\ %
6 & SK [5.5,6.5] M1                   &$-0.41$&$[-0.63,\,-0.29]$&$-0.27$&$-0.40$&$-0.14$\\ %
7 & SK [5.5,6.5] M2                   &$-1.77$&$[-2.00,\,-1.64]$&$-1.84$&$-1.72$&$-1.43$\\ %
8 & SK [5.5,6.5] M3                   &$ 0.03$&$[-0.20,\,+0.26]$&$ 0.05$&$ 0.19$&$ 0.52$\\ %
9 & SK [5.5,6.5] M4                   &$ 1.11$&$[+0.93,\,+1.36]$&$ 1.46$&$ 1.30$&$ 1.64$\\ %
10& SK [5.5,6.5] M5                   &$ 0.62$&$[+0.47,\,+0.86]$&$ 0.71$&$ 0.80$&$ 1.14$\\ %
11& SK [5.5,6.5] core                 &$-0.90$&$[-1.04,\,-0.67]$&$-0.75$&$-0.73$&$-0.41$\\ %
12& SK [6.5,8.0] day                  &$ 1.20$&$[+0.48,\,+2.35]$&$ 1.15$&$ 0.56$&$ 0.72$\\ %
13& SK [6.5,8.0] M1                   &$ 1.97$&$[+1.80,\,+2.04]$&$ 1.99$&$ 1.84$&$ 1.90$\\ %
14& SK [6.5,8.0] M2                   &$ 1.31$&$[+0.82,\,+1.38]$&$ 1.08$&$ 1.29$&$ 1.37$\\ %
15& SK [6.5,8.0] M3                   &$-1.29$&$[-1.80,\,-1.07]$&$-1.39$&$-1.13$&$-1.02$\\ %
16& SK [6.5,8.0] M4                   &$ 0.19$&$[-0.23,\,+0.46]$&$ 0.62$&$ 0.41$&$ 0.55$\\ %
17& SK [6.5,8.0] M5                   &$-0.81$&$[-1.17,\,-0.54]$&$-0.78$&$-0.60$&$-0.40$\\ %
18& SK [6.5,8.0] core                 &$-1.12$&$[-1.41,\,-0.87]$&$-0.99$&$-0.94$&$-0.73$\\ %
19& SK [8.0,9.5] day                  &$-0.40$&$[-0.93,\,+0.56]$&$-0.67$&$-0.91$&$-1.29$\\ %
20& SK [8.0,9.5] M1                   &$ 0.17$&$[+0.05,\,+0.19]$&$ 0.13$&$ 0.09$&$-0.05$\\ %
21& SK [8.0,9.5] M2                   &$ 0.24$&$[-0.32,\,+0.42]$&$ 0.06$&$ 0.31$&$ 0.17$\\ %
22& SK [8.0,9.5] M3                   &$-0.17$&$[-0.73,\,+0.15]$&$-0.22$&$ 0.08$&$-0.05$\\ %
23& SK [8.0,9.5] M4                   &$ 1.40$&$[+0.92,\,+1.76]$&$ 1.75$&$ 1.70$&$ 1.61$\\ %
24& SK [8.0,9.5] M5                   &$-0.26$&$[-0.65,\,+0.09]$&$-0.20$&$ 0.02$&$ 0.03$\\ %
25& SK [8.0,9.5] core                 &$-0.51$&$[-0.83,\,-0.21]$&$-0.39$&$-0.28$&$-0.23$\\ %
26& SK [9.5,11.5] day                 &$-0.67$&$[-1.19,\,+0.19]$&$-1.07$&$-0.83$&$-1.70$\\ %
27& SK [9.5,11.5] M1                  &$-0.20$&$[-0.36,\,-0.13]$&$-0.29$&$-0.17$&$-0.49$\\ %
28& SK [9.5,11.5] M2                  &$ 1.01$&$[+0.45,\,+1.30]$&$ 0.88$&$ 1.22$&$ 0.90$\\ %
29& SK [9.5,11.5] M3                  &$-0.55$&$[-1.11,\,-0.10]$&$-0.56$&$-0.13$&$-0.49$\\ %
30& SK [9.5,11.5] M4                  &$ 0.08$&$[-0.43,\,+0.57]$&$ 0.40$&$ 0.55$&$ 0.26$\\ %
31& SK [9.5,11.5] M5                  &$ 0.17$&$[-0.21,\,+0.63]$&$ 0.26$&$ 0.57$&$ 0.45$\\ %
32& SK [9.5,11.5] core                &$ 0.33$&$[+0.01,\,+0.69]$&$ 0.43$&$ 0.62$&$ 0.52$\\ %
33& SK [11.5,13.5] day                &$ 0.72$&$[+0.36,\,+1.31]$&$ 0.42$&$ 0.66$&$ 0.11$\\ %
34& SK [11.5,13.5] M1                 &$ 0.67$&$[+0.57,\,+0.71]$&$ 0.61$&$ 0.71$&$ 0.52$\\ %
35& SK [11.5,13.5] M2                 &$-2.21$&$[-2.64,\,-1.98]$&$-2.25$&$-2.00$&$-2.23$\\ %
36& SK [11.5,13.5] M3                 &$-1.76$&$[-2.11,\,-1.43]$&$-1.72$&$-1.39$&$-1.65$\\ %
37& SK [11.5,13.5] M4                 &$ 0.36$&$[+0.02,\,+0.69]$&$ 0.57$&$ 0.70$&$ 0.55$\\ %
38& SK [11.5,13.5] M5                 &$-0.80$&$[-1.02,\,-0.50]$&$-0.72$&$-0.59$&$-0.54$\\ %
39& SK [11.5,13.5] core               &$ 0.67$&$[+0.46,\,+0.92]$&$ 0.76$&$ 0.76$&$ 0.82$\\ %
40& SK [13.5,16.0] day                &$ 0.65$&$[+0.36,\,+1.02]$&$ 0.50$&$-0.37$&$ 0.48$\\ %
41& SK [13.5,16.0] M1                 &$ 0.59$&$[+0.52,\,+0.61]$&$ 0.56$&$ 0.25$&$ 0.57$\\ %
42& SK [13.5,16.0] M2                 &$ 2.03$&$[+1.88,\,+2.06]$&$ 2.03$&$ 1.79$&$ 2.08$\\ %
43& SK [13.5,16.0] M3                 &$ 1.51$&$[+1.39,\,+1.57]$&$ 1.54$&$ 1.26$&$ 1.62$\\ %
44& SK [13.5,16.0] M4                 &$ 1.31$&$[+1.18,\,+1.39]$&$ 1.42$&$ 1.05$&$ 1.47$\\ %
45& SK [13.5,16.0] M5                 &$-0.48$&$[-0.57,\,-0.41]$&$-0.42$&$-0.85$&$-0.26$\\ %
46& SK [13.5,16.0] core               &$-0.22$&$[-0.30,\,-0.15]$&$-0.15$&$-0.60$&$-0.08$\\ %
47& SK [16.0,20.0]                    &$ 0.34$&$[+0.06,\,+0.47]$&$ 0.31$&$-0.19$&$ 0.35$\\ %
48& SNO [5.0,5.5] day                 &$-0.99$&$[-1.32,\,-0.28]$&$-0.49$&$-0.66$&$-0.08$\\ %
49& SNO [5.0,5.5] night               &$ 0.23$&$[-0.17,\,+1.01]$&$ 0.78$&$ 0.65$&$ 1.19$\\ %
50& SNO [5.5,6.0] day                 &$ 0.94$&$[+0.63,\,+1.42]$&$ 1.23$&$ 1.06$&$ 1.33$\\ %
51& SNO [5.5,6.0] night               &$-0.31$&$[-0.65,\,+0.12]$&$-0.36$&$-0.64$&$-0.60$\\ %
52& SNO [6.0,6.5] day                 &$ 0.09$&$[-0.53,\,+0.64]$&$-0.26$&$-0.60$&$-0.74$\\ %
53& SNO [6.0,6.5] night               &$-0.72$&$[-1.55,\,-0.02]$&$-1.27$&$-1.52$&$-1.88$\\ %
54& SNO [6.5,7.0] day                 &$-0.64$&$[-1.46,\,+0.10]$&$-1.24$&$-1.20$&$-1.82$\\ %
55& SNO [6.5,7.0] night               &$ 1.34$&$[+0.70,\,+1.95]$&$ 0.84$&$ 1.17$&$ 0.40$\\ %
56& SNO [7.0,7.5] day                 &$-0.66$&$[-1.24,\,-0.05]$&$-1.14$&$-0.57$&$-1.51$\\ %
57& SNO [7.0,7.5] night               &$ 0.27$&$[-0.13,\,+0.74]$&$-0.09$&$ 0.38$&$-0.31$\\ %
58& SNO [7.5,8.0] day                 &$ 0.41$&$[+0.15,\,+0.79]$&$ 0.16$&$ 0.37$&$ 0.06$\\ %
59& SNO [7.5,8.0] night               &$ 0.39$&$[+0.25,\,+0.68]$&$ 0.24$&$ 0.13$&$ 0.22$\\ %
60& SNO [8.0,8.5] day                 &$-0.20$&$[-0.26,\,+0.04]$&$-0.27$&$-0.65$&$-0.21$\\ %
61& SNO [8.0,8.5] night               &$-0.69$&$[-0.73,\,-0.47]$&$-0.68$&$-1.20$&$-0.58$\\ %
62& SNO [8.5,9.0] day                 &$-1.31$&$[-1.39,\,-1.07]$&$-1.25$&$-1.78$&$-1.10$\\ %
63& SNO [8.5,9.0] night               &$ 0.43$&$[+0.35,\,+0.60]$&$ 0.51$&$ 0.27$&$ 0.60$\\ %
64& SNO [9.0,9.5] day                 &$-0.72$&$[-0.96,\,-0.39]$&$-0.51$&$-0.20$&$-0.44$\\ %
65& SNO [9.0,9.5] night               &$ 0.54$&$[+0.07,\,+1.35]$&$ 1.03$&$ 1.09$&$ 1.71$\\ %
66& SNO [9.5,10.0] day                &$-0.95$&$[-1.59,\,+0.07]$&$-0.29$&$-0.19$&$ 0.50$\\ %
67& SNO [9.5,10.0] night              &$-0.66$&$[-1.13,\,-0.07]$&$-0.27$&$-0.21$&$ 0.17$\\ %
68& SNO [10.0,10.5] day               &$ 0.76$&$[+0.44,\,+0.85]$&$ 0.80$&$ 0.75$&$ 0.84$\\ %
69& SNO [10.0,10.5] night             &$ 2.19$&$[+1.70,\,+2.43]$&$ 1.96$&$ 1.84$&$ 1.75$\\ %
70& SNO [10.5,11.0] day               &$ 0.60$&$[-0.09,\,+1.02]$&$ 0.20$&$ 0.18$&$-0.14$\\ %
71& SNO [10.5,11.0] night             &$-1.76$&$[-2.54,\,-1.22]$&$-2.24$&$-1.93$&$-2.60$\\ %
72& SNO [11.0,11.5] day               &$ 0.78$&$[+0.15,\,+1.24]$&$ 0.41$&$ 0.99$&$ 0.17$\\ %
73& SNO [11.0,11.5] night             &$-0.69$&$[-1.25,\,-0.21]$&$-0.98$&$-0.21$&$-1.13$\\ %
74& SNO [11.5,12.0] day               &$ 0.27$&$[-0.17,\,+0.69]$&$ 0.10$&$ 0.69$&$ 0.08$\\ %
75& SNO [11.5,12.0] night             &$-0.27$&$[-0.64,\,+0.10]$&$-0.33$&$-0.08$&$-0.25$\\ %
76& SNO [12.0,12.5] day               &$ 1.28$&$[+1.01,\,+1.54]$&$ 1.31$&$ 1.17$&$ 1.45$\\ %
77& SNO [12.0,12.5] night             &$ 0.89$&$[+0.66,\,+1.14]$&$ 1.00$&$ 0.59$&$ 1.19$\\ %
78& SNO [12.5,13.0] day               &$-0.45$&$[-0.69,\,-0.13]$&$-0.26$&$-0.84$&$-0.03$\\ %
79& SNO [12.5,13.0] night             &$ 0.64$&$[+0.44,\,+0.92]$&$ 0.82$&$ 0.40$&$ 1.02$\\ %
80& SNO [13.0,13.5] day               &$ 1.37$&$[+1.21,\,+1.60]$&$ 1.54$&$ 1.33$&$ 1.68$\\ %
81& SNO [13.0,13.5] night             &$-1.03$&$[-1.41,\,-0.57]$&$-0.64$&$-0.19$&$-0.51$\\ %
\hline\hline
\end{longtable}
\end{table}
\endgroup

Table~\ref{obspulls} shows the pulls of each of the 81 observables
used in our analysis, corresponding to the four solutions in
Table~\ref{chisquares}.%
\footnote{The SK and SNO spectrum bins are identified by their
energy range and by their nadir interval (day and night bins). See
also Apps.~\ref{skspectrum} and \ref{snotreat} for notation.}
For the best-fit LMA solution, we also give the range spanned by
each pull within the $99.73\%$ C.L.\ LMA region shown in
Fig.~\ref{fig01}. It can be seen that, in any case, there are no
anomalously large (say, $>3$ standard deviation) pulls---not even
in the SMA solution or at the borders of the LMA region. This fact
confirms the results of the previous sections, namely, that the
allowance or rejection of the various solutions comes from a
``collective'' effect of several experimental observables rather
than by a small subset of them.

\begin{figure}
\includegraphics[scale=0.96, bb= 100 100 510 750]{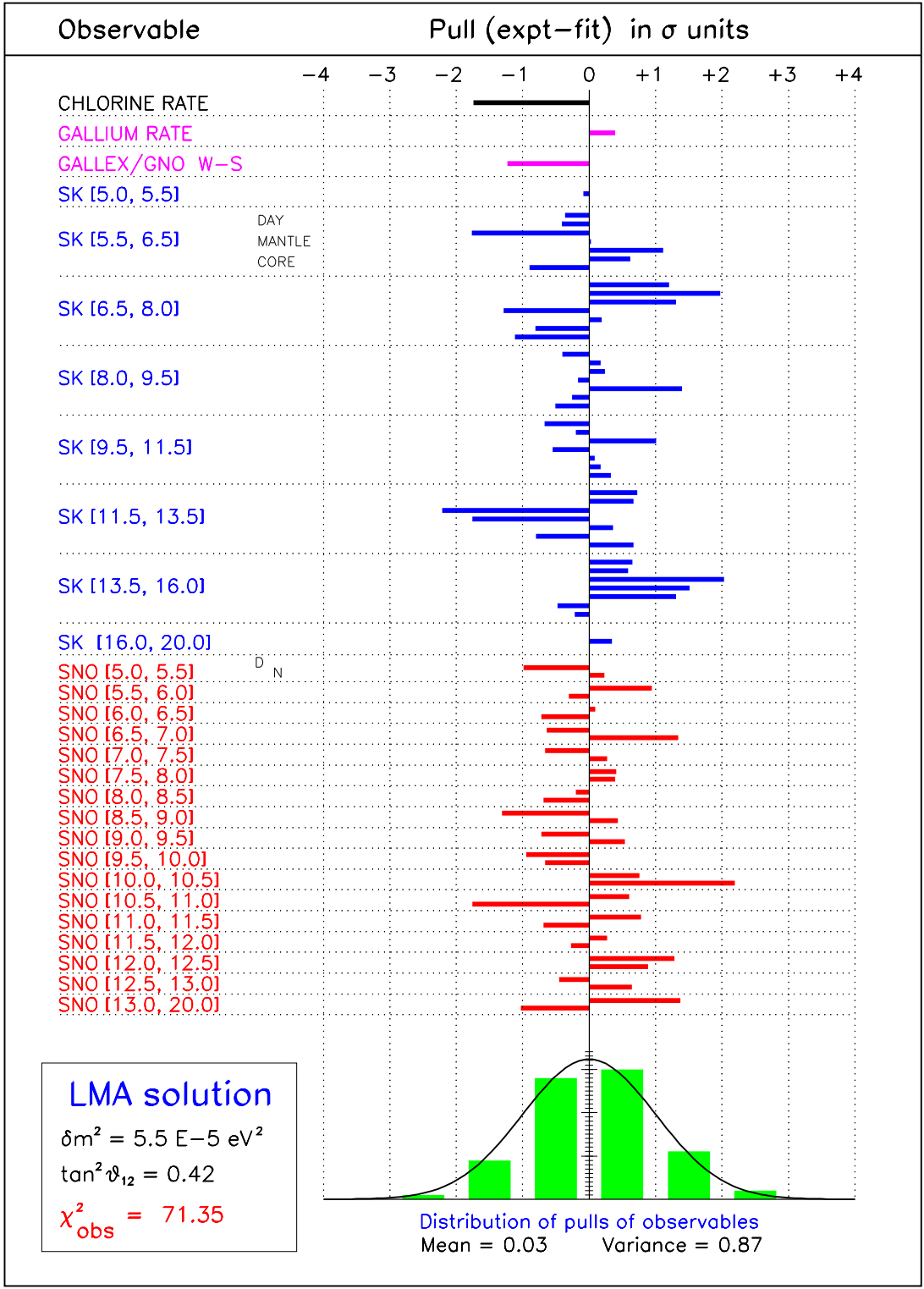}
\vspace*{-0.4cm} \caption{ \label{fig04} Diagram of pulls
$\{\overline x_n\}_{n=1,\dots,81}$ for observables  at the LMA
best-fit point. See the text for details.}
\end{figure}

Figure~\ref{fig04} provides a graphical version of the first three
columns in Table~\ref{obspulls}, together with an histogram of the
pull distribution, compared with a univariate gaussian
distribution (conventionally normalized to $N=81$) to guide the
eye. It appears that the pulls of the observables have a
reasonably symmetrical and gaussian distribution, confirming the
goodness of the LMA fit at a deeper level than the global $\chi^2$
values. The relatively large pull for the Chlorine datum ($-1.74$)
may be regarded as a statistical fluctuation among the others, at
the LMA best fit. Notice that the Cl pull in the SMA solution is
instead very small (0.14). Retrospectively, this small pull,
together with theoretical prejudices against large mixing, appears
to be at the origin of a very long detour towards small-mixing
oscillations in matter. The distributions of pulls of the
observables for the QVO and LOW cases (not graphically shown) are
also reasonably gaussian as for the LMA solution, although with a
slightly larger area (given by the corresponding
$\chi^2_\mathrm{obs}$ values in Table~\ref{chisquares}). The
distribution of pulls for the SMA case (not shown), besides having
an even larger area ($\chi^2_\mathrm{obs}=83.1$), appears also to
be slightly skewed, with an excess of positive pulls. This adds to
the statistical problems of this solution.

It is interesting to group the separate experimental contributions
to the global $\chi^2_\mathrm{obs}$ in Table~\ref{chisquares}, by
summing up the corresponding squared pulls from
Table~\ref{obspulls}. The results of this exercise, as reported in
Table~\ref{groupobs}, show which experiment ``wins'' or ``looses''
in the various global solutions. In particular, the radiochemical
experiments clearly win in the SMA and loose in the QVO and LOW
solutions, while the fit to the SK spectrum observables appears to
be rather stable, with only $\Delta\chi^2_\mathrm{obs}\simeq \pm 3
$ variations in the various solutions. The SMA solution tries to
make a compromise between SK and SNO data, in which SK dominates
(having smaller spectral errors), leaving SNO with a worse fit
($\chi^2_\mathrm{obs}=38.5$) as compared with the LMA case
($\chi^2_\mathrm{obs}=26.2$).

\begingroup
\squeezetable
\begin{table}[t]
\caption{\label{groupobs} \footnotesize\baselineskip=4mm Separate
experimental contributions to the global $\chi^2_\mathrm{obs}$ for
the various solutions reported in Table~\ref{chisquares}, as
obtained by grouping squared pulls from Table~\ref{obspulls}.}
\begin{ruledtabular}
\begin{tabular}{clrrrr}
$n$ & Experiment                & LMA   & LOW   & QVO   & SMA \\
\hline
1       & Cl (rate)             & 3.0   & 3.4   & 7.2   & 0.0\\
2--3    & Ga (rate and $W-S$)   & 1.7   & 4.7   & 2.3   & 1.5\\
4--48   & SK (44  bins)         & 40.4  & 42.9  & 37.0  & 43.1\\
49--81  & SNO (34 bins)        & 26.2  & 28.7  & 28.4  & 38.5\\
\hline
1--81& All (global $\chi^2_\mathrm{obs})$ &71.3&79.7&74.9  & 83.1
\end{tabular}
\end{ruledtabular}
\end{table}
\endgroup

\begingroup \squeezetable
\begin{table}[b]
\begin{longtable}{@{\extracolsep{39.5pt}}clrcrrr}
\caption{\label{syspulls} Pulls of the systematics in the various solutions.}\\[-2mm]
\hline\hline\\
k & Systematic &LMA& LMA [min,~max] at $3\sigma$ & LOW & QVO & SMA\\ %
\hline
1 & $S_{11}$                &$-0.05$&$[-0.31,\,+0.42]$&$ 0.19$&$ 0.35$&$ 0.34$\\ %
2 & $S_{33}$                &$ 0.00$&$[-0.13,\,+0.32]$&$ 0.10$&$ 0.24$&$ 0.18$\\ %
3 & $S_{34}$                &$ 0.01$&$[-0.98,\,+0.42]$&$-0.32$&$-0.72$&$-0.59$\\ %
4 & $S_{1,14}$              &$-0.15$&$[-0.49,\,-0.02]$&$-0.08$&$-0.21$&$-0.01$\\ %
5 & $S_{17}$                &$ 0.38$&$[-0.71,\,+1.15]$&$-0.35$&$-0.47$&$-0.83$\\ %
6 & Luminosity              &$ 0.04$&$[-0.31,\,+0.17]$&$-0.08$&$-0.20$&$-0.22$\\ %
7 & $Z/X$                   &$ 0.03$&$[-0.89,\,+0.48]$&$-0.35$&$-0.68$&$-0.60$\\ %
8 & Age                     &$ 0.00$&$[-0.06,\,+0.04]$&$-0.02$&$-0.05$&$-0.04$\\ %
9 & Opacity                 &$-0.05$&$[-0.36,\,+0.52]$&$ 0.22$&$ 0.42$&$ 0.41$\\ %
10& Diffusion               &$-0.02$&$[-0.26,\,+0.43]$&$ 0.18$&$ 0.34$&$ 0.31$\\ %
11& $C_\mathrm{Be}$         &$-0.07$&$[-0.22,\,+0.13]$&$ 0.07$&$ 0.09$&$ 0.16$\\ %
12& $S_{\rm hep}$           &$-0.03$&$[-0.04,\,+0.00]$&$-0.02$&$-0.10$&$ 0.02$\\ %
13& $^8$B $\nu$ shape       &$ 0.17$&$[-0.71,\,+1.24]$&$-0.66$&$-0.80$&$-1.56$\\ %
14& SK scale                &$ 0.78$&$[+0.51,\,+1.82]$&$ 0.49$&$ 0.49$&$-0.31$\\ %
15& SK resolution           &$ 0.61$&$[+0.54,\,+0.87]$&$ 0.61$&$ 0.06$&$ 0.73$\\ %
16& SK offset               &$ 0.44$&$[+0.33,\,+0.70]$&$ 0.57$&$ 0.68$&$ 0.34$\\ %
17& SK [5.0,5.5]            &$-0.03$&$[-0.18,\,+0.05]$&$ 0.06$&$ 0.00$&$ 0.27$\\ %
18& SK [5.5,6.5]            &$-0.26$&$[-0.61,\,-0.13]$&$-0.10$&$-0.28$&$ 0.34$\\ %
19& SK [6.5,8.0]            &$ 0.54$&$[+0.32,\,+0.67]$&$ 0.70$&$ 0.52$&$ 0.89$\\ %
20& SK [8.0,9.5]            &$ 0.01$&$[-0.06,\,+0.19]$&$-0.08$&$-0.03$&$-0.42$\\ %
21& SK [9.5,11.5]           &$-0.14$&$[-0.25,\,+0.21]$&$-0.30$&$ 0.14$&$-0.76$\\ %
22& SK [11.5,13.5]          &$-0.21$&$[-0.31,\,-0.06]$&$-0.29$&$-0.10$&$-0.45$\\ %
23& SK [13.5,16.0]          &$ 0.26$&$[+0.23,\,+0.34]$&$ 0.32$&$ 0.11$&$ 0.44$\\ %
24& SK [16.0,20.0]          &$ 0.01$&$[+0.00,\,+0.02]$&$ 0.02$&$-0.01$&$ 0.02$\\ %
25& SNO scale               &$-0.15$&$[-0.90,\,+0.58]$&$-0.86$&$-1.49$&$-1.48$\\ %
26& SNO resolution          &$-0.32$&$[-0.41,\,-0.05]$&$-0.16$&$-0.52$&$ 0.47$\\ %
27& SNO vertex              &$ 0.13$&$[-0.60,\,+0.65]$&$-0.52$&$-0.20$&$-1.42$\\ %
28& SNO $n$ capture         &$-0.10$&$[-0.46,\,+0.60]$&$ 0.42$&$ 0.34$&$ 0.94$\\ %
29& SNO $n$ background      &$-0.06$&$[-0.27,\,+0.35]$&$ 0.25$&$ 0.20$&$ 0.55$\\ %
30& SNO LE background       &$-0.16$&$[-0.49,\,+0.53]$&$ 0.33$&$ 0.28$&$ 0.87$\\ %
31& SNO cross section       &$ 0.04$&$[-0.16,\,+0.21]$&$-0.16$&$-0.02$&$-0.52$\\ %
\hline\hline
\end{longtable}
\end{table}
\endgroup

\begin{figure}
\includegraphics[scale=0.90, bb= 100 100 510 780]{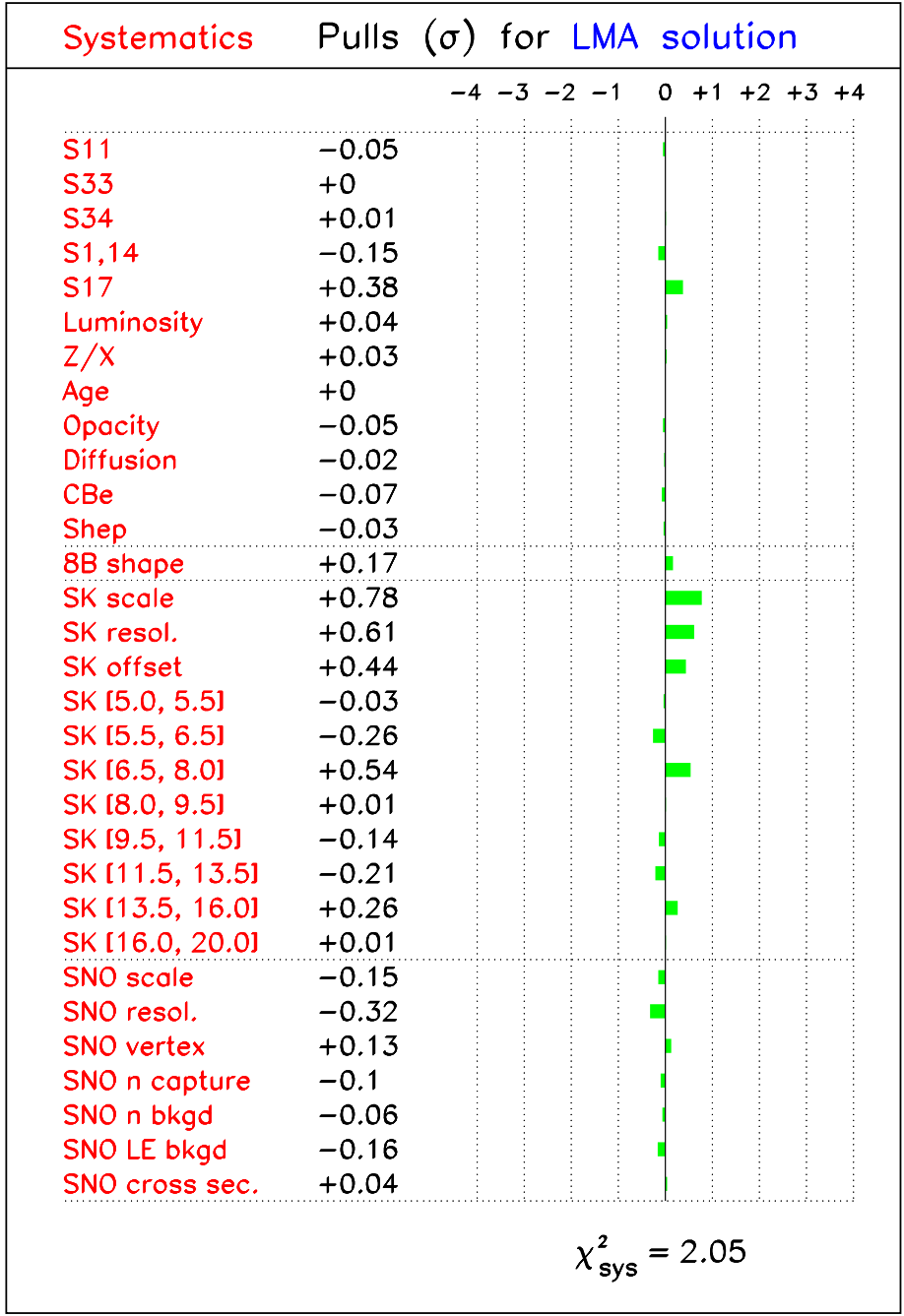}
\vspace*{-1cm} \caption{\footnotesize \label{fig05} Diagram of
pulls $\{\overline \xi_k\}_{k=1,\dots,31}$ for correlated
systematics at the LMA best-fit point. See the text for details.}
\end{figure}

Let us now consider the contributions of correlated systematics
shifts to the global fit. Table~\ref{syspulls} shows the
contributions of the pulls of systematics in the various
solutions. As in Table~\ref{obssys}, for the LMA solution we also
show the range spanned by the pulls within the $99.73\%$ C.L.\ LMA
region in Fig.~\ref{fig01}. It appears that such pulls are
generally rather small, and typically assume minimal values in the
LMA solution. In particular, Fig.~\ref{fig05} shows the LMA pull
diagram for the correlated systematics, none of which presents a
significant offset. The smallness of all SSM input offsets (with
the possible exception of $S_{17}$) is particularly impressive. In
general, at the LMA best-fit point there is no need to stretch any
correlated systematics to fit the data. In the LOW and QVO
solutions the situation is less ideal (several offsets in
Table~\ref{syspulls} are at the level of $\sim 0.5\sigma$) but
certainly still acceptable. In the SMA case, however, several SNO
systematic offsets in Table~\ref{syspulls} are at the level of
$\sim 1\sigma$, and the $^8$B $\nu$ shape uncertainty is stretched
by $\sim 1.5\sigma$. Such offsets, which act in the direction of
reducing the spectral differences between theory and data, produce
a significant contribution to the SMA fit
($\chi^2_\mathrm{sys}=13.8)$. It is quite unlikely that future,
possible recalibrations of systematics may just happen to cancel
out all these offsets, thus giving more chances to the SMA
solution.

It is interesting to group some $\chi^2$ contributions of
systematics, according to their origin. This exercise is done in
Table~\ref{groupsys}. The SK contribution to $\chi^2_\mathrm{sys}$
is rather stable (and small), with a minimum at the QVO solution.
This situation parallels the SK contribution to
$\chi^2_\mathrm{obs}$ (see Table~\ref{groupobs} and related
comments), and show the pivoting role of the SK spectrum in
determining the likelihood of all solutions. Conversely, the
contributions to $\chi^2_\mathrm{sys}$ from SNO and from the
standard neutrino flux input (SSM and $^8$B $\nu$ shape
systematics) increase significantly when passing from the LMA to
the SMA solution.

\begingroup
\squeezetable
\begin{table}[t]
\caption{\label{groupsys} \footnotesize\baselineskip=4mm Separate
contributions to the global $\chi^2_\mathrm{sys}$ for the various
solutions reported in Table~\ref{chisquares}, as obtained by
grouping squared pulls from Table~\ref{syspulls}.}
\begin{ruledtabular}
\begin{tabular}{clrrrr}
$k$ & Systematic\ sources               & LMA   & LOW   & QVO   & SMA \\
\hline
1--13& SSM and $^8$B $\nu$ shape&       0.2  &  1.0   & 2.4   & 4.3\\
14--24 & SK systematics          &       1.7   & 1.7   & 1.1   & 2.9\\
25--31 & SNO systematics          &      0.2  &  1.4  &  2.8  &  6.6\\
\hline
1--31& All (global $\chi^2_\mathrm{sys})$ &2.1&4.1&6.3  & 13.8
\end{tabular}
\end{ruledtabular}
\end{table}
\endgroup

The grand total of the various contributions to the
$\chi^2_\mathrm{pull}$ from both observables and systematics is
shown in Table~\ref{groupall}. This Table shows quantitatively
that the LMA solution is in very good agreement with both the
experimental data and with the SSM. The LOW and QVO solutions
provide a slightly less good agreement with the SSM and with
SK+SNO, and are somehow ``borderline'' from the point of view of
radiochemical experiments. However, they cannot be really excluded
by any data at present. The SMA solution is instead safely ruled
out, mainly as a consequence of the SNO data fit
($\chi^2_\mathrm{SNO}=45.1$).

\begingroup
\squeezetable
\begin{table}[h]
\caption{\label{groupall} \footnotesize\baselineskip=4mm Separate
contributions to the global $\chi^2_\mathrm{pull}$ for the various
solutions reported in Table~\ref{chisquares}.}
\begin{ruledtabular}
\begin{tabular}{cclrrrr}
$n$ & $k$ &       Contributions               & LMA   & LOW   & QVO   & SMA \\
\hline
 ---   & 1--13 & SSM and $^8$B $\nu$ shape & 0.2  &  1.0   & 2.4   & 4.3\\
 1      &   --- & Cl experiment             & 3.0  &  3.4   & 7.2   & 0.0\\
 2--3   &   --- & Ga experiments            & 1.7  &  4.7   & 2.3   & 1.5\\
 4--48  &14--24 & SK experiment             &42.1  & 44.6   &38.1   &46.0\\
 49--81 &25--31 & SNO experiment            &26.4  & 30.1   &31.2   &45.1\\
\hline
1--81& 1--31 & All (global $\chi^2_\mathrm{})$ &73.4&83.8&81.2  & 96.9\\
\end{tabular}
\end{ruledtabular}
\end{table}
\endgroup

\subsection{Implications for the SSM neutrino fluxes}

In the previous section, we have seen that the LMA solution does
not require any significant offset $\overline\xi_k$ in the SSM
input systematics ($k=1,\dots,12$), and that such offsets are
relatively small also in the other solutions.
Equation~\ref{shiftflux} allows to translate such offsets into
preferred shift of the neutrino fluxes $\Phi_i$ from their SSM
central values. The results are shown in Table~\ref{fluxes} for
the various solutions. In the LMA best-fit point, all solar
neutrino fluxes are basically confirmed: just for illustration, an
approximate translation of the LMA flux shifts into variations of
the ``effective solar core temperature'' ($\Delta t_c/t_c$
\cite{Ul96,Ca97,Li00}) would formally provide $|\Delta
t_c/t_c|\lesssim 0.2\%$. The QVO and SMA solutions would formally
require $-\Delta t_C/t_C\sim 0.5$--1\% (a slightly cooler Sun),
the LOW case being intermediate between the latter and the LMA
one. In all solutions, the preferred $\Phi_\mathrm{pp}$ values are
within a percent from the SSM. The preferred (negative) variations
of the $\Phi_B$ values from its SSM value are instead significant
in non-LMA solutions, and appear to be consistent with those
derived in some $\Phi_B$-free analyses \cite{Pe02,Ho02,Smy0}. This
might seem surprising, since we do not exclude any SSM input in
our analysis. The reason for such agreement is that the current
SNO data implicitly provide an experimental determination of
$\Phi_B$ which is already significantly more accurate than the SSM
estimate; therefore, there is little difference in making the
analysis with or without SSM input for $\Phi_B$. Notice that, in
addition,  we can also quantify the preferred variations of
$\Phi_\mathrm{Be}$, $\Phi_N$, and $\Phi_O$, which
appear to be all negative in non-LMA solutions.%
\footnote{This information is relevant for prospective studies in
the BOREXINO experiment \cite{BORE}.}
Concerning $\Phi_\mathrm{hep}$, we do not find any significant
preferred variation with respect to Eq.~(\ref{hepflux}). This fact
is mainly due to the poor sensitivity of the data to this flux
(confined to the last few bins in the SK and SNO spectrum) and
possibly to our careful treatment of SK and SNO spectral
uncertainties (see Apps.~\ref{skspectrum} and \ref{snotreat}).

\begin{table}[t]
\caption{\label{fluxes} Fractional neutrino flux shifts from the
SSM central values ($\Delta\Phi_i/\Phi_i\times 100$), for the
various solutions.}
\begin{ruledtabular}
\begin{tabular}{rrrrrrrr}
        & pp   & pep    & Be    & B     & hep   & N     & O \\
\hline

LMA     & 0.0  & 0.0    & 0.5   & 5.2   &$-0.8$ &$-1.0$ &$-1.2$\\
LOW     & 0.5  & 0.8    &$-5.5$ &$-12.2$&$0.0$  &$-8.3$ &$-8.3$\\
QVO     & 1.1  & 1.6    &$-11.5$&$-22.2$&$-1.8$ &$-15.1$&$-17.0$\\
SMA     & 0.9  & 1.3    &$-9.9$ &$-24.2$&$+1.8$ &$-11.4$&$-12.8$\\
\end{tabular}
\end{ruledtabular}
\end{table}

\begin{figure}
\includegraphics[scale=0.90, bb= 100 100 510 760]{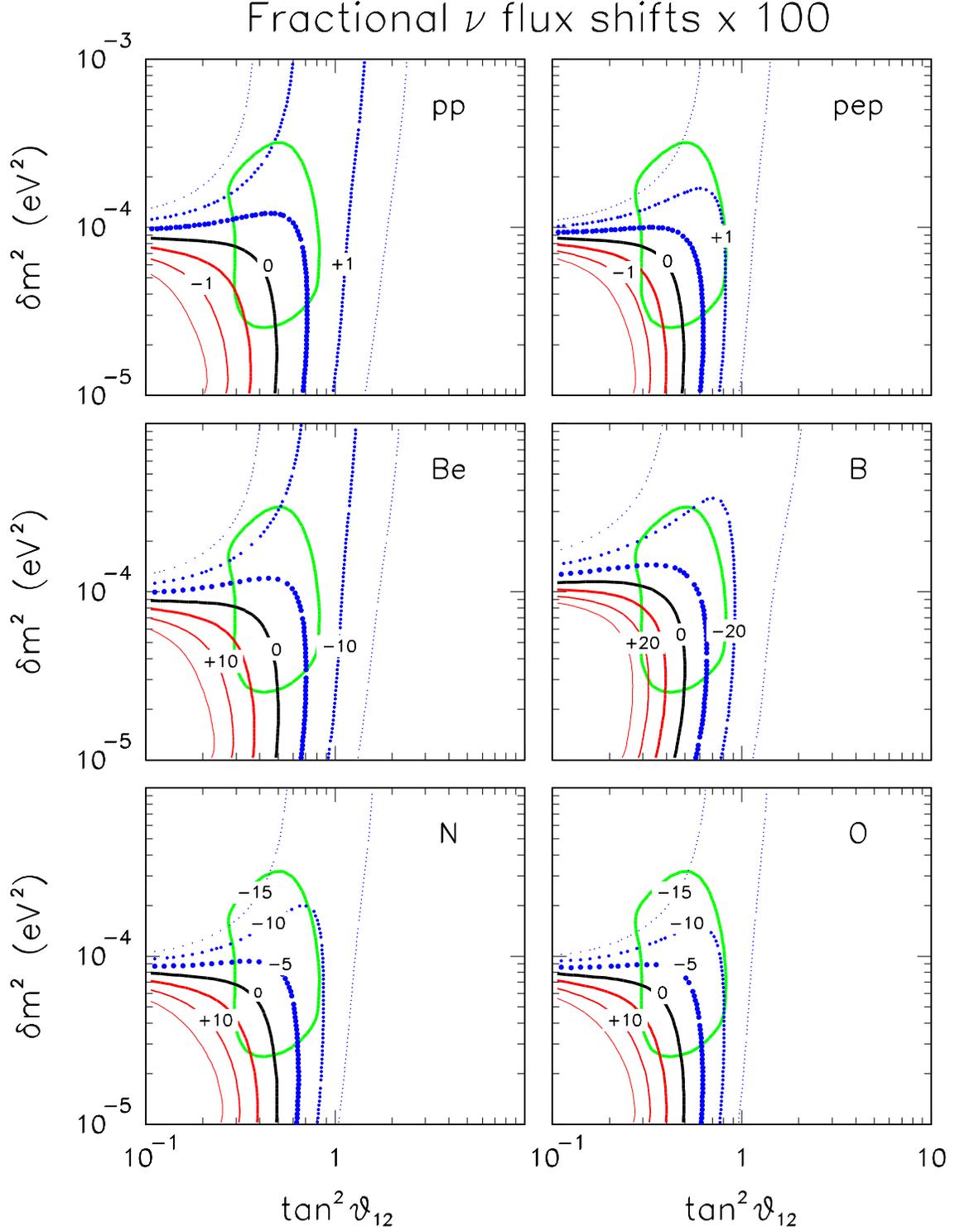}
\vspace*{+0.3cm} \caption{\label{fig06} Preferred deviations of
the solar $\nu$ fluxes from their SSM central values \cite{Ba01},
in the region of the LMA solution (superposed at 99\% C.L.).
Dotted (solid) isolines roughly correspond to a slightly
``cooler'' (hotter) Sun. See the text for details.}
\end{figure}

Finally, Fig.~\ref{fig06} shows isolines of the preferred neutrino
flux shifts within the LMA solution (superposed at 99\% C.L.). The
variations of $\Phi_\mathrm{pp}$ and of $\Phi_\mathrm{pep}$ are
limited to about a percent, but can be an order of magnitude
larger for the other neutrino fluxes. In the upper or right part
of the LMA region, the current global fits seem to prefer
variations of the fluxes roughly corresponding to a slightly
``cooler'' Sun, while in the lower left corner of the LMA region
the trend is opposite (slightly ``hotter'' Sun). Should the
best-fit LMA value be confirmed by the KamLand experiment
\cite{KamL}, the current SSM input would be just ``perfect.''

\section{Summary and Conclusions}

We have performed a global analysis of solar neutrino oscillations
within the $2\nu$ active scenario,%
\footnote{Three-flavor results will be shown elsewhere.}
including all the current solar neutrino data and all relevant
sources of uncertainties. The statistical analysis has been
performed in a way which clearly traces the residual contribution
of each observable and of each source of correlated systematics in
the $\chi^2$ fit (``pull'' method). It turns out that there is
still a multiplicity of acceptable solutions at large mixing in
the so-called LMA, LOW, and QVO regions (Fig.~\ref{fig01} and
Table~\ref{chisquares}), none of which contradict any data
(Figs.~\ref{fig02} and \ref{fig03}), as also confirmed by a
detailed pull analysis (Tables~\ref{obspulls}--\ref{groupall}). In
particular, the best-fit LMA solution appears to be in very good
agreement with all the data (Fig.~\ref{fig04}), without requiring
any stretching of the correlated systematics (Fig.~\ref{fig05}),
contrary to the strongly disfavored solution at small mixing
angle. The striking LMA agreement with all standard solar model
fluxes (Table~\ref{fluxes}) is only slightly worsened when moving
away from the best fit (Fig.~\ref{fig06}). The statistical
techniques and the treatment of the data underlying these results
have been discussed in detail in various appendices.

\acknowledgments

We thank the organizers of the {\em 20th International Conference
on Neutrino Physics and Astrophysics\/} in Munich---where
preliminary results of this work were presented---for kind
hospitality. We also thank the many participants to this
Conference who have contributed useful comments, remarks, and
suggestions. This work was supported in part by INFN and in part
by the Italian {\em Ministero dell'Istruzione, Universit\`a e
Ricerca\/} through the ``Astroparticle Physics'' research project.

\appendix

\section{Proof of $\chi^2_\mathrm{covar}=\chi^2_\mathrm{pull}$}
\label{theorem}

Let us remind that basic ingredients of any $\chi^2$ statistics
are: the experimental and theoretical values ($R_n^\mathrm{expt}$
and $R_n^\mathrm{theor}$) of the $N$ observables to be fitted; the
associated uncorrelated errors $u_n$; and the associated set of
fully correlated errors $c_n^k$, due to $K$ independent sources of
systematics ($k=1,\dots,K$).

In order to simplify the notation, we normalize both the
differences $R_n^\mathrm{expt}-R_n^\mathrm{theo}$ and the
correlated errors $c_n^k$ to the $u_n$'s, by defining
\begin{equation}\label{Deltan}
\Delta_n = \frac{R_n^\mathrm{expt}-R_n^\mathrm{theor}}{u_n} \ ,
\end{equation}
and
\begin{equation}\label{qnk}
q^k_n=\frac{c^k_n}{u_n}\ .
\end{equation}

Equation~(\ref{chicovardef}) reads then
\begin{equation}\label{chicovardef2}
\chi^2_\mathrm{covar} = \sum_{n,m=1}^N \Delta_n
[\delta_{nm}+{\scriptstyle\sum_k} q_n^k q_m^k]^{-1} \Delta_m\ ,
\end{equation}
while Eq.~(\ref{chipulldef}) reads
\begin{equation}\label{chipulldef2}
\chi^2_\mathrm{pull} = \min_{\{\xi_k\}}\left[\sum_{n=1}^N
(\Delta_n-{\scriptstyle\sum_k} q_n^k\, \xi_k)^2 + \sum_{k=1}^K
\xi^2_k\right]\ ,
\end{equation}
where the $\xi_k$ are gaussian random variables with $\langle
\xi_k\rangle =0$ and $\langle \xi_k^2\rangle=1$.

The minimization in Eq.~(\ref{chipulldef2}) leads to a set of $K$
linear equations in the unknowns $\overline\xi_k$,
\begin{equation}\label{minequation}
\sum_{h=1}^K (\delta_{kh}+{\scriptstyle \sum_n}q_n^k
q_n^h)\overline\xi_h=\sum_{n=1}^N \Delta_n q^k_n\ ,
\end{equation}
whose solution is
\begin{equation}\label{csibar}
  \overline\xi_k=\sum_{h=1}^K S_{kh} \sum_{n=1}^N\Delta_n q^h_n\ ,
\end{equation}
where $S$ is the inverse matrix%
\footnote{Here we deal only with symmetric matrices. Therefore,
given a matrix equation such as $\mathbf{A}=\mathbf{B}^{-1}$, we
can conventionally write it as $A_{nm}=[B_{nm}]^{-1}$ without
index ambiguity. }
\begin{equation}\label{Skh}
S_{kh}=[\delta_{kh}+{\scriptstyle \sum_n}q_n^k q_n^h]^{-1}\ .
\end{equation}

It turns out that the  matrix $S$ is also related to the inversion
of the  covariance matrix,
\begin{equation}\label{variance}
[\delta_{nm}+{\scriptstyle\sum_k} q_n^k q_m^k]^{-1} =
\delta_{nm}-\sum_{k,h=1}^K S_{kh}\, q_m^h\,q_n^k\ .
\end{equation}
Indeed, the product of the above two matrices gives the unit
matrix. The above equation reduces the inversion of the $N\times
N$ covariance matrix in $\chi^2_\mathrm{covar}$ to the inversion
of the (generally much smaller) $K\times K$ matrix in
Eq.~(\ref{Skh}).

The last step is to write the $\chi^2_\mathrm{pull}$ in terms of
the $\overline\xi_k$'s,
\begin{equation}\label{chipulldef3}
\chi^2_\mathrm{pull} = \sum_{n=1}^N (\Delta_n-{\scriptstyle\sum_k}
q_n^k\, \overline\xi_k)^2 + \sum_{k=1}^K \overline\xi^2_k\ ,
\end{equation}
and to substitute Eq.~(\ref{csibar}) in Eq.~(\ref{chipulldef3}).
Expanding the r.h.s.\ of Eq.~(\ref{chipulldef3}), and making use
of Eq.~(\ref{variance}), one recovers the r.h.s\ of
Eq.~(\ref{chicovardef2}), namely,
\begin{equation}\label{finalproof}
\chi^2_\mathrm{covar}\equiv\chi^2_\mathrm{pull} \ .
\end{equation}

Finally, we observe that setting
\begin{equation}\label{xbarnapp}
\overline x_n \equiv \Delta_n-\sum_{k=1}^K q_n^k\overline\xi_k\ ,
\end{equation}
as in Eq.~(\ref{xbarn}), one gets from Eq.~(\ref{chipulldef3}) a
``diagonal'' form for $\chi^2_\mathrm{pull}$,
\begin{equation}\label{pulldiagapp}
\chi^2_\mathrm{pull}=\sum_{n=1}^N \overline x^2_n +
\sum_{k=1}^K\overline \xi_k^2\ ,
\end{equation}
as anticipated in Eq.~(\ref{chipulldiag}).

A different proof of the previous relations have been discussed in
the context of parton distribution fitting \cite{Stum}, where the
pull method is now routinely used \cite{Bo02,Pu02}.

\section{Winter-Summer asymmetry in GALLEX/GNO}
\label{wsapp}

Earth matter effects can generate an observable winter-summer
difference ($R_W-R_S$) in the event rates measured in gallium
experiments. Such difference can be as large as $\sim6$ SNU around
the LOW solution \cite{Fo00}.

The GALLEX/GNO Collaboration has recently reported the measurement
\cite{Ki02} (see also \cite{TAUP})
\begin{equation}
\label{wsgno} R_W-R_S=-11\pm 9\mathrm{\ SNU\ (GALLEX/GNO)}\ ,
\end{equation}
whose uncertainty is almost entirely statistical \cite{TAUP}, all
systematics being largely cancelled in the difference \cite{Catt}.
In our ``pull'' approach, we simply attach to $R_W-R_S$ an
uncorrelated error $u_\mathrm{WS}=9$~SNU, with no systematics.

\begin{figure}
\includegraphics[scale=0.90, bb= 100 100 510 720]{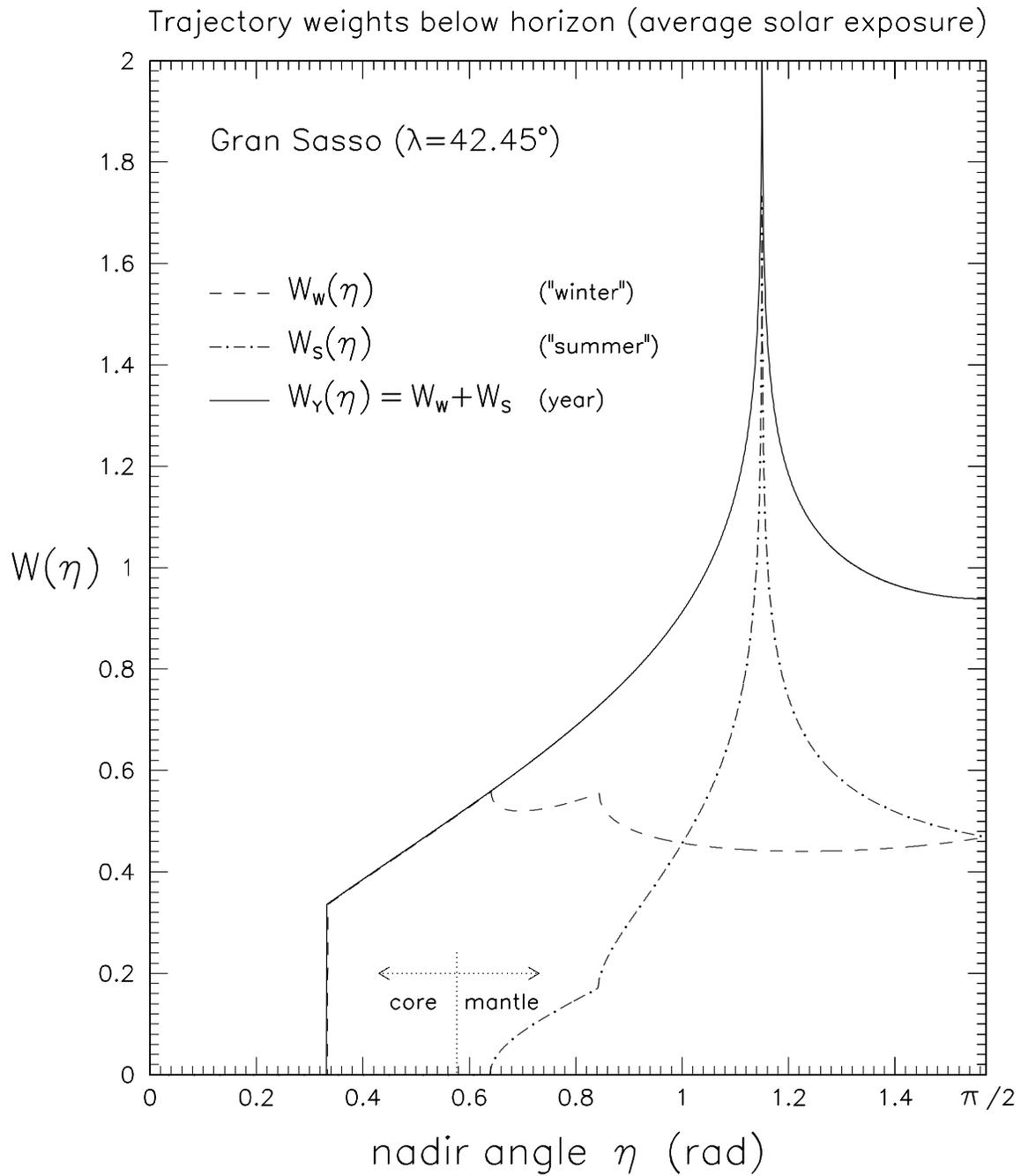}
\vspace*{+0cm} \caption{ \label{fig07} Solar exposure functions
(at the Gran Sasso latitude) for the ``winter'' and ``summer''
periods defined in Eqs.~(\ref{wdef}) and (\ref{sdef}).}
\end{figure}

The definition of ``Winter'' and ``Summer'' adopted in
\cite{Ki02,TAUP} is slightly
different from the astronomical one,%
\footnote{Winter and summer were defined in \cite{Fo00} as
six-months intervals separated by equinoxes and centered,
respectively, at the winter solstice (21 dec.) and at the summer
solstice (21 jun).}
and corresponds to
\begin{eqnarray}
\label{wdef} \mathrm{``Winter"} &=& \mathrm{perihelion\ (5\
jan.)}\pm 3\mathrm{\ months}\ ,\\
\label{sdef} \mathrm{``Summer"} &=& \mathrm{aphelion\  \ \ (5\
july)}\pm 3\mathrm{\ months}\ .
\end{eqnarray}
Therefore, the solar exposure functions for the above periods, as
shown in Fig.~\ref{fig07} in terms of the nadir angle ($\eta$),
are slightly different from the ones given in \cite{Fo00}. In
particular, the comparison of Fig.~\ref{fig07} in this paper with
Fig.~5 in \cite{Fo00} shows the the above definitions lead to a
lower (higher) exposure of the innermost trajectories in the
mantle during winter (summer). The total annual exposure is, of
course, unaltered.

The definitions in Eqs.~(\ref{wdef}) and (\ref{sdef}) are
particularly useful to smoothly extend the theoretical calculation
of $R_W-R_S$ from the matter-enhanced regime \cite{Fo00}  down to
the vacuum oscillation regime, where seasonal variations are
instead induced by the eccentricity variations of the orbital
distance $L$ (purely geometrical ($1/L^2$) effects being factored
out). Indeed, with the definitions in Eqs.~(\ref{wdef}) and
(\ref{sdef}), the winter-summer rate difference in vacuum happens
to coincide with the near-far rate difference ($R_N-R_F$)
previously defined in \cite{Faid}. We have then matched the
results found in \cite{Fo00} and in \cite{Faid}, by considering
both matter-induced and eccentricity-induced contributions to
$R_W-R_S$, so as to calculate this quantity in the {\em whole\/}
oscillation parameter space. The match between the matter and
vacuum regimes occurs in the quasivacuum range, and is made easier
by the lucky circumstance that Earth matter effects vanish just
when the oscillating terms in the $\nu_e$ survival probability
start to be important \cite{QVOs}.

The datum in Eq.~(\ref{wsgno}) is compatible with no seasonal
asymmetry, adds a slight penalty to the region roughly
corresponding to the LOW solution (where $0\lesssim
R_W-R_S\lesssim 6$~SNU), and modulates the likelihood of the
solutions in the Q(VO) regime, where both positive and negative
values of the asymmetry can occur ($-26 \lesssim R_W-R_S\lesssim
+26$~SNU), the negative ones being slightly favored by the
GALLEX/GNO datum. We remark that (positive or negative) seasonal
effects in gallium experiments, being largely driven by low-energy
solar neutrino components, can be compatible with the
nonobservation of ($^8$B-driven) seasonal effects in SK
\cite{Smy0}.

Finally, we mention that the SAGE experiment has  recently
reported $\nu$ event rates in gallium, grouped in (bi)monthly
intervals \cite{Ab02}, which appear to be consistent with no
seasonal variations (although no explicit $R_W-R_S$ estimate is
given in \cite{Ab02}). Taking the data in \cite{Ab02} at face
value, we argue a slightly positive value for $R_W-R_S$ in SAGE,
with a total uncertainty comparable to that of GALLEX/GNO. The
combination of a slightly positive (SAGE) and a slightly negative
(GALLEX/GNO) winter-summer difference would then provide a central
value closer to zero for $R_W-R_S$, very consistent with the LMA
solution \cite{Fo00}. An official evaluation of $R_W-R_S$ from
SAGE [to be combined with the one in Eq.~(\ref{wsgno}) from
GALLEX/GNO] appears thus a desirable input for future analyses.

\section{The SK energy-nadir differential spectrum}
\label{skspectrum}

The SK energy-nadir spectrum of electrons induced by neutrino
elastic scattering  is fundamental to constrain the solar neutrino
parameter space. Therefore, we think it useful to describe in some
detail our improved error estimates and theoretical calculations.

Concerning the ``error format'', the pull method used in this
paper does not leave any freedom in the $\chi^2$ treatment of the
spectrum, which is uniquely defined by providing, for each bin
rate $R_n$, the uncorrelated error component $u_n$, and the fully
correlated error components $c^k_n$ due to independent $k$-th
sources of systematic
errors.%
\footnote{Therefore, we abandon our previous $\chi^2$ approach in
terms of separated SK total rate and spectral shape information
\cite{Sudb} which, although correct, cannot be exactly cast in a
``pull'' form.}
The task is thus reduced to a careful evaluation of such
components. To reach this goal, we combine information from SK
\cite{Fu02} and from our own evaluation of systematics.

Table~\ref{skspect} shows the main characteristics of the SK
binned energy spectrum. The energy bins 2--7 are further divided
into seven nadir angle bins \cite{Fu02}, reported in
Table~\ref{sknadir}. In each energy-nadir bin, the statistical
error represents the only SK uncorrelated error component ($u_n$).
All other error sources in SK are correlated among bins, at least
in nadir. Indeed, apart from the obvious SSM systematics, the SK
bin rates are further affected by $8+3+1=12$ correlated
systematics, as discussed in the following.

\begingroup
\squeezetable
\begin{table}[t]
\caption{\label{skspect} \footnotesize \baselineskip=4.mm
Characteristics of the SK energy spectrum. The first two columns
define the bins in terms of observed total electron energy $E$.
The third column is the ratio of the SK observed event rates to
the SSM \cite{Ba01} predictions, as taken from \cite{Fu02}, with a
slight correction to account for our default hep $\nu$ flux [we
take $\Phi_\mathrm{hep}$=8.3  rather than 9.3 \cite{Ba01} in units
of $\times 10^3$ cm$^{-2}$~s$^{-1}$, see Eq.~(\ref{hepflux})]. The
boron flux is $\Phi_B=5.05\times 10^6$ cm$^{-2}$~s$^{-1}$
\cite{Ba01}, as in \cite{Fu02}. The fourth column gives the
fractional uncorrelated error in each bin, whose only component is
the statistical error (taken from \cite{Fu02}). The fifth column
represents those fractional systematic errors which are
uncorrelated among energy bins, but correlated in nadir bins (from
\cite{Fu02}). The sixth, seventh and eighth columns represent our
evaluation of the total, boron, and hep $\nu$ event rate in SK (in
the absence of oscillation), for 1~kton~year (kty) exposure. The
ninth, tenth, and eleventh column represent our evaluation of the
three main systematic effects (given as $1\sigma$ fractional
contributions to $R_{m,B}^\mathrm{theo}$) which are fully
correlated in energy and in nadir, and are generated by
uncertainties in the $^8$B $\nu$ energy shape, SK energy scale,
and SK resolution width. The last two uncertainties also affect
the hep contribution to the total rate, as reported in the last
two columns (our evaluation, given as $1\sigma$ fractional
contributions to $R_{m,{\rm hep}}^\mathrm{theo}$). Finally, we
attach an overall systematic offset $c^\mathrm{off}\simeq 2.75 \%$
\cite{Fu02} to  all binned rates $R_m^\mathrm{theo}$, with full
correlation in energy and nadir. In the presence of oscillations,
the last eight columns (properly separated into nadir bins) are
recalculated for each $(\delta m^2,\tan^2\theta_{12})$ point.}
\begin{ruledtabular}
\begin{tabular}{ccccc|rrr|crr|rr} Bin & $E$ range &
$R^\mathrm{expt}_m/R^\mathrm{theor}_m$ & $u_m/R^\mathrm{expt}_m$ &
$c_m^\mathrm{bin}/R^\mathrm{theor}_m$ &
$R_{m}^\mathrm{theor}$\footnotemark[1] & $R_{m,B}^\mathrm{theor}$&
$R_{m,\mathrm{hep}}^\mathrm{theor}$ &
\multicolumn{3}{c|}{$100\times c_{m,B}/R_{m,B}^\mathrm{theor}$} &
\multicolumn{2}{c}{$100\times c_{m,\mathrm{hep}}/R_{m,\mathrm{hep}}^\mathrm{theor}$}\\
$m$ & (MeV) & (B+hep)& $\times 100$ & $\times 100$ & (1/kty) &
(1/kty) & (1/kty) & $^8$B shape & scale & resol. & scale & resol.
\\
\hline
1&$[5.0,5.5]$  &0.4672&8.65&3.24& 81.795& 81.624&0.171&0.52&$-0.07$&$-0.19$&$-0.38$&$-0.09$ \\
2&$[5.5,6.5]$  &0.4581&3.08&1.43&140.974&140.654&0.320&0.69& 0.14  &$-0.20$&$-0.29$&$-0.11$   \\
3&$[6.5,8.0]$  &0.4730&1.78&1.37&155.433&155.014&0.419&1.05& 0.58  &$-0.20$&$-0.09$&$-0.15$  \\
4&$[8.0,9.5]$  &0.4601&2.02&1.37& 95.577& 95.239&0.338&1.61& 1.32  &$-0.09$&  0.24 &$-0.18$ \\
5&$[9.5,11.5]$ &0.4630&2.23&1.37& 60.011& 59.688&0.323&2.48& 2.50  &0.00   &  0.78 &$-0.19$\\
6&$[11.5,13.5]$&0.4626&3.64&1.37& 18.197& 18.004&0.193&3.87& 4.50  &1.59   &  1.67 &$-0.07$\\
7&$[13.5,16.0]$&0.5683&6.88&1.37&  3.768&  3.660&0.108&5.82& 7.47  &4.51   &  3.14 &0.42\\
8&$[16.5,20.0]$&0.5637&26.3&1.37&  0.245&  0.212&0.033&8.09&12.32
&11.26  &  6.09 &2.41
\end{tabular}
\end{ruledtabular}
\footnotetext[1]{\scriptsize The total event rate
$R^\mathrm{theo}$ is here normalized to the efficiency-corrected,
no-oscillation value of 556 events/kty, as graphically derived
from \cite{Fuku}, with a correction for the updated SSM
$\Phi_{B,\rm{hep}}$ values. This specific value for
$R^\mathrm{theo}$ is unimportant in practice, as far as the ratios
in the 3rd column are used in the analysis.}
\end{table}
\endgroup

The first eight systematic errors, listed in the fifth column of
Table~\ref{skspect}, are uncorrelated among energy bins. In each
energy bin, however, they are fully correlated in nadir
\cite{Fu02}. Such errors represent those (nadir-independent) data
reduction uncertainties which are specific of each energy bin,
independently of the others.

Three further systematics, fully correlated among all energy-nadir
bins, are induced by  $^8$B $\nu$ spectrum shape uncertainties
\cite{Boro} and by the SK energy scale and resolution
uncertainties \cite{Fu02}. We note two facts. Firstly, these
sources of systematics act differently on the B and hep components
in SK: for instance, the
hep component is obviously unaffected by the $^8$B shape uncertainty.%
\footnote{Analyses where the $\Phi_\mathrm{hep}/\Phi_B$ ratio is
taken as a free parameter should separately rescale the different
$\Phi_B$ and $\Phi_\mathrm{hep}$ error components in the
evaluation of the total error. In fact, one cannot calculate first
the total ($\Phi_B+\Phi_\mathrm{hep}$) systematic error assuming
the SSM ratio for $\Phi_\mathrm{hep}/\Phi_B$, and then use the
same  error when such ratio is significantly varied (say, by
factors $\sim 10$). We think that this remark should be taken into
account, when placing upper bounds on $\Phi_\mathrm{hep}$ from the
analysis of the high-energy tail of the SK spectrum.}
Secondly, the relative error signs (for each source of
systematics) are relevant: for instance, an increase in the energy
resolution width flattens the---steeply falling---SK energy
spectrum, the high-energy (low-energy) part being thus enhanced
(suppressed), as evident from the eleventh and thirteenth column
in Table~\ref{skspect}. Since the information about the relative
error signs and about the separate B and hep components is not
given in \cite{Fu02}, we perform our own evaluation as follows.

The $^8$B spectrum shape error is evaluated by attaching to the
default neutrino spectrum \cite{Or00} the $\pm 1\sigma$ shape
perturbations evaluated in \cite{Boro}, and calculating the
corresponding
fractional variations in the absolute rates.%
\footnote{Conventionally, we denote as $+1\sigma$ perturbation of
the  $^8$B $\nu$ shape the one which moves the energy spectrum to
higher energies.}
The energy scale uncertainty is evaluated by shifting the centroid
of the (gaussian) energy resolution function by 0.64\%
\cite{Fu02}, namely, by taking $T'\to T'(1+0.0064)$, where $T'$ is
the true electron kinetic energy (the quantity calibrated in SK).
The resolution uncertainty is evaluated by perturbing the energy
resolution width $\sigma_T$ (see Eq.~\ref{sigmask}) by 2.5\%
\cite{Fu02}, $\sigma_T\to\sigma_T(1+0.025)$. In all cases, the
same uncertainties are calculated for opposite perturbations, and
small error asymmetries are averaged. The results are given in the
last five columns of Table~\ref{skspect}, where the contributions
from $\Phi_B$ and $\Phi_\mathrm{hep}$ are separated. As far as the
B flux component is concerned, there is reasonable agreement in
size with the corresponding SK error evaluation in \cite{Fu02},
except for the $^8$B $\nu$ spectrum shape. We are unable to
explain such difference. Our error assignment is completed by an
overall SK systematic offset ($2.75\%$, symmetrized from
\cite{Fu02}), which is attached to all energy-nadir theoretical
rates with full correlation. This error mainly represents the
overall uncertainty of the data reduction efficiency \cite{Smy0},
affecting the whole spectrum.

\begingroup
\squeezetable
\begin{table}[t]
\caption{\label{sknadir} \footnotesize \baselineskip=4.mm
Fractional contributions to the total event rate
($R/R_\mathrm{tot}$, our evaluation), in the case of no
oscillation, from each of the seven SK nadir intervals, identified
as ``day'' bin, ``mantle'' bins (M1--M5), and ``core'' bin. The
observed SK event rates in each energy-nadir bin, together with
their statistical errors, are reported in \cite{Fu02}.}
\begin{ruledtabular}
\begin{tabular}{lcr}
Nadir bin & $\cos\eta$\ range & $R/R_\mathrm{tot}$\\
\hline
Day  &$[-1,0]$ &0.5000  \\
M1   &$[0,0.16]$ &0.0685  \\
M2   &$[0.16,0.33]$ &0.0777  \\
M3   &$[0.33,0.50]$ &0.0984  \\
M4   &$[0.50,0.67]$ &0.1025  \\
M5   &$[0.67,0.84]$ &0.0839  \\
Core &$[0.84,1]$ &0.0690  \\
\hline
Day+Night &$[-1,1]$& 1.0000
\end{tabular}
\end{ruledtabular}
\end{table}
\endgroup

Notice that, in Table~\ref{skspect}, the last eight columns refer
to the no oscillation case. In the presence of oscillations, one
usually updates only the theoretical rates $R_m^\mathrm{theo}$,
and assumes that the fractional errors in the last five columns
are approximately unchanged. The latter assumption is a very good
approximation for spectra with no or mild distortions, but is not
strictly applicable in the whole oscillation parameter space.
Therefore, for the sake of accuracy, in our oscillation analysis
we recalculate the fractional errors in the last five columns of
Table~\ref{skspect}, for any $(\delta m^2,\tan^2\theta_{12})$ grid
point (and for any nadir bin). This improvement leads to minimal
differences in the local $\chi^2_\mathrm{pull}$ minima
(corresponding to almost undistorted spectra), but is not totally
negligible in deriving the {\em borders\/} of, say, the 99.73\%
C.L.\  allowed regions, where it can lead to $|\Delta\chi^2|\simeq
1$ differences.

Concerning the calculations of the $\nu_e$ survival probability,
we compute Earth matter effects \cite{Eart} through eight relevant
shells of the Preliminary Reference Earth Model (PREM)
\cite{PREM}, namely (in the language of \cite{PREM}): (1) Ocean;
(2,3) Crust layers; (4) LID + low velocity zone; (5) Transition
zone; (6) Lower Mantle + D-zone; (7) Outer core; and (8) Inner
core. In each of them, we approximate the radial density profile
through a biquadratic parameterization, which allows a fast and
accurate analytical calculation of the relevant transition
probabilities in the Earth \cite{Eart}.

\begin{figure}
\includegraphics[scale=0.85, bb= 100 100 510 790]{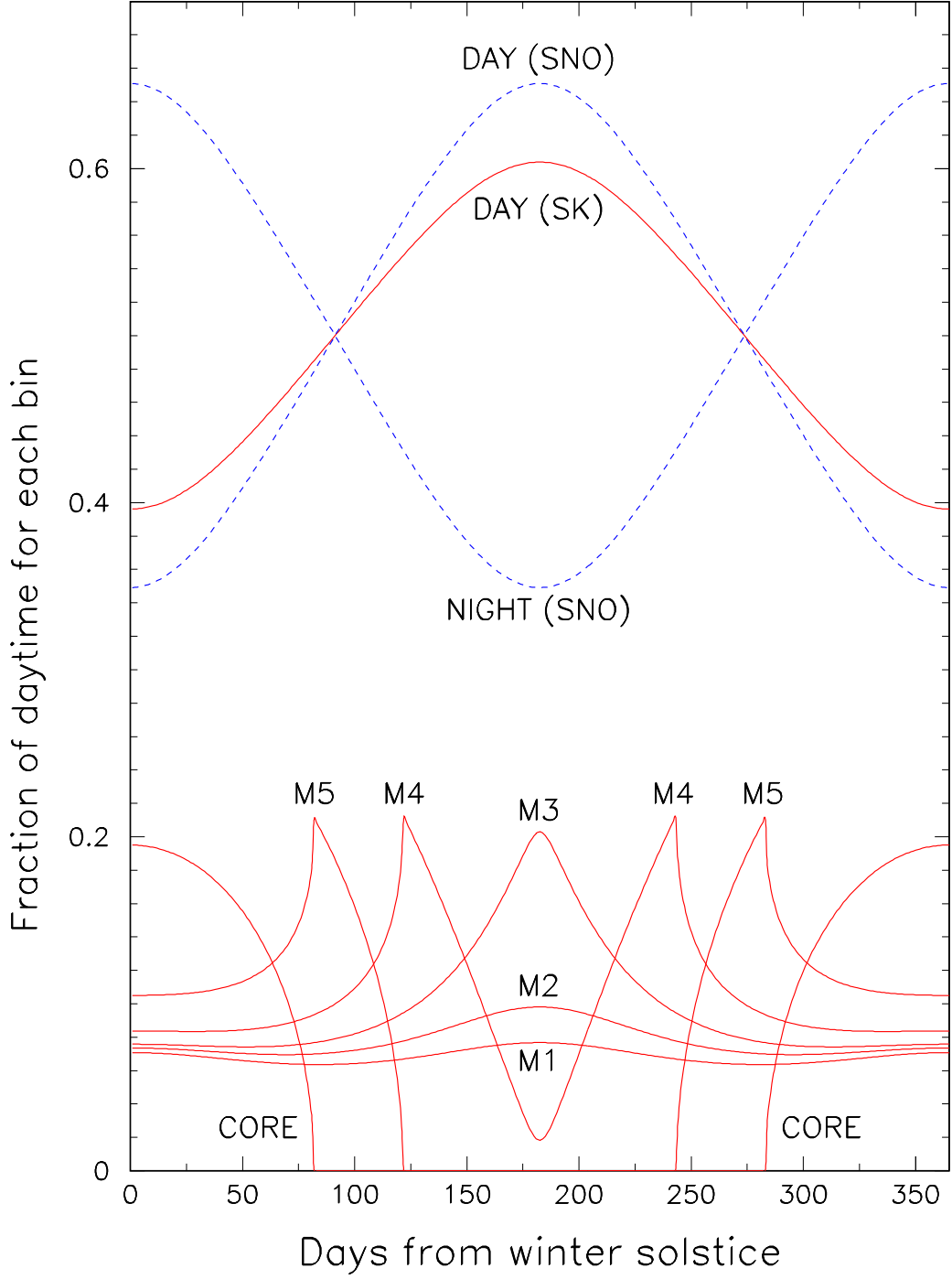}
\vspace*{+0cm} \caption{ \label{fig08} Daytime solar exposures of
each SK nadir bin and of the SNO day and night bins. These
exposure functions are used for accurate time averages of the
oscillating terms in the (quasi)vacuum regime.}
\end{figure}

In the (quasi)vacuum oscillation regime, the oscillating term in
the $\nu_e$ survival probability $P_{ee}$ depends implicitly upon
the daily time through the orbital distance $L(\tau_d)$, where
$\tau_d = 2\pi\cdot
\mathrm{day}/365$,  with $\tau_d=0$ at winter solstice.%
\footnote{The purely geometrical $1/L^2(\tau_d)$ variation is
assumed to be already corrected for in the SK data.}
As noticed in \cite{VaDN}, the different exposure of the ``day''
and ``night'' bins in terms of $\tau_d$ induces a slight day-night
{\em vacuum\/} difference in the time-averaged $P_{ee}$.
Concerning the SK energy-nadir spectrum, we take into account the
different exposures $E(\eta_1,\eta_2)$ for each nadir bin range
$[\eta_1,\eta_2]$ in Table~\ref{sknadir} as follows. In a given
day of the year ($\tau_d$), the $\eta$ range spanned by a detector
at latitude $\lambda$ (equal to $36.48^\circ$ for SK) is
$[\eta_{\min}(\tau_d),\eta_{\max}(\tau_d)]$, where
$\eta_\mathrm{min}(\tau_d)=\lambda +\delta_S(\tau_d)$, and
$\eta_\mathrm{max}(\tau_d)=\pi-\lambda +\delta_S(\tau_d)$, having
defined the solar declination $\delta_S(\tau_d)$ through
$\sin\delta_S=-\sin i \cos\tau_d$, with $\sin i=0.3978$
(inclination of Earth axis). The daily exposure function $E$ for a
generic $[\eta_1,\eta_2]$ nadir bin is then
\begin{equation}\label{Eeta}
E(\tau_d,\eta_1,\eta_2)=\left[\tau_h(\tau_d,\min\{\eta_2,\eta_{\max}\})
- \tau_h(\tau_d,\max\{\eta_1,\eta_{\min}\})\right]/\pi\ ,
\end{equation}
where $\tau_h(\tau_d,\eta)$ is the hourly time (normalized to
$2\pi$ and starting at midnight) corresponding to the angle $\eta$
during the day $\tau_d$,%
\footnote{Any angle $\eta\in[\eta_{\min},\eta_{\max}]$ is spanned
twice each day. This explains the appearance of $\pi$ (rather than
of $2\pi$) in the denominator of Eq.~(\ref{Eeta}).}
\begin{equation}
\label{hours}
\tau_h(\tau_d,\eta)=\arccos\left(\frac{\cos\eta}{\cos\lambda\cos\delta_S}
+\tan\lambda\tan\delta_S \right)\ .
\end{equation}
Figure~\ref{fig08} shows the exposure functions $E(\tau_d)$ for
the seven SK nadir bins listed in Table~\ref{sknadir}, as well as
for the SNO day and night bins (the functions sum up to unity in
both cases). It can be seen, as intuitively expected, that the SK
core bin is sensitive only to extreme values of $L(\tau_d)$ (close
to the orbital perihelion and aphelion), while the two outermost
mantle bins (M1 and M2) are almost equally sensitive to all values
of $L(\tau_d)$, being crossed by solar neutrinos during the whole
year. We take into account the different bin exposures in
Fig.~\ref{fig08} when time-averaging the oscillating terms in the
(quasi)vacuum regime, for both SK and SNO.

The results of our oscillation analysis of the SK spectrum are
given in the SK panel of the (previously discussed)
Fig.~\ref{fig02}. The C.L.\ contours compare well with the
official SK ones, as reported in Fig.~2 of \cite{Fu02} (we also
obtain very good agreement for the normalization-free case, not
shown).

A final remark is in order. All the SK systematic error sources
discussed in \cite{Fu02} (and adopted in this work) are fully
correlated in nadir. This means that no allowance is given in
\cite{Fu02} for uncorrelated nadir shape variations, apart from
the obvious statistical fluctuations. However, it is known from
previous SK publications \cite{Fu01,DayN} and presentations
\cite{Fuku} that some systematics {\em do not\/} cancel in
day-night differences, and may thus, in general, affect the nadir
bins in uncorrelated ways. In addition, Earth matter density and
chemical composition uncertainties may also contribute to small
(and different) errors \cite{Brig,Yerr} in the various nadir bins
\cite{Sh01} via $\nu_e$ regeneration effects. In particular, one
should not forget that the widely used PREM model is spherically
symmetric by construction, and that local departures from the
world-averaged density are to be expected. For instance, the first
half of the SK mantle bin M1 (see Table~\ref{sknadir}) is
sensitive to the PREM ``Ocean+Crust'' density, whose local
characteristics at the SK site may well be different from the
world average. The inclusion of such additional uncertainties
might give a little more freedom to fit possible distortions in
the nadir distributions. Therefore, if the (currently weak) hints
for an excess of night events in SK \cite{Smy0} and in SNO
\cite{AhDN} will be corroborated by future data, an improved
discussion of the nadir spectrum uncertainties will be useful to
precisely assess their statistical significance.

\section{Treatment of the SNO day-night energy spectrum}
\label{snotreat}

The solar neutrino events observed in the SNO spectrum \cite{AhNC}
cannot be currently identified as being of ES, CC, or NC type on
an event-by-event basis. A statistical separation is possible,
however, by exploiting their different distribution in terms of
suitable variables, the most important being the observed
(effective) electron kinetic energy $T$. Since the ES and CC
distribution shapes in $T$ depend upon the oscillation parameters,
also the inferred CC and NC rates depend on such parameters, as
stressed in \cite{HOWT}.

In the oscillation analysis, however, it is not necessary to
perform a separate fit to the ES, CC, and NC components as done in
\cite{Pe02}. Given the oscillated predictions for these
components, one can simply add them up (together with the known
background rates) in each energy bin, calculate the corresponding
systematics, and fit the observed SNO day-night energy spectrum.
This method, which has been dubbed ``forward fitting'' by the SNO
Collaboration \cite{HOWT}, allows to take into account the full
spectral information (central values and errors of each bin). In
the following, we describe our implementation of such method in
the $\chi^2_\mathrm{pull}$ evaluation.

Table~\ref{snospect} reports some relevant characteristics of the
SNO spectrum, including our evaluation of the SNO neutrino signal
components and their errors (small error asymmetries being
averaged out). The effects of the $^8$B $\nu$ spectrum shape error
(not included in the official SNO analysis \cite{AhNC,AhDN}) are
estimated as in SK (see App.~\ref{skspectrum}). Notice that a
$+1\sigma$ shift of this systematic uncertainty increases the
number of signal events, especially at high energies. The
fractional increase is obviously constant for the NC events, which
have no memory of the original $\nu$ energy. From
Table~\ref{snospect} it appears that, in the bulk of the SNO
spectrum, the $^8$B $\nu$ shape uncertainties are not negligible
as compared with the (purely instrumental) SNO energy calibration
and resolution uncertainties. The latter two errors are evaluated
by shifting the centroid of the energy resolution function and
varying its width in the way described in Ref.~\cite{HOWT}, and
calculating the corresponding variations in each bin of the CC,
NC, and ES spectrum. Actually we separate the $\Phi_B$ and
$\Phi_\mathrm{hep}$ components (not shown in
Table~\ref{snospect}), as for the SK spectrum. In the presence of
oscillations, we not only update the day and night rates in each
bin, but also their {\em fractional\/} systematic errors, for each
point in the mass-mixing plane. As for SK, the effect of this
improved estimate of fractional systematic errors is rather small
in the local $\chi^2$ minima, but increases towards the borders of
the 99.73\% C.L.\ regions, where the spectral distortions can be
more sizeable. In the fit to SNO data only, such improvement can
lead to $\chi^2_\mathrm{pull}$ variations as large as $|\Delta
\chi^2|\sim 4$ at the $3\sigma$ borders, and is thus not totally
negligible in deriving precise C.L.\ contours for the less favored
solutions. Finally, notice that the three systematic error sources
in Table~\ref{snospect} can induce, in the various bins, event
rate variations with different relative signs, which are taken
into account in the analysis.

Besides the previous three systematic error sources, we include a
vertex reconstruction uncertainty, whose $+1\sigma$ effect is to
increase the CC and ES rates by $+ 3\%$ and the NC rates by
$+1.45\%$ \cite{HOWT}. The CC and NC binned rates are also
affected by a systematic cross section uncertainty, whose
$+1\sigma$ effect is to increase them by $+1.8\%$ and $+1.3\%$,
respectively \cite{AhNC}.%
\footnote{The correlation between the NC and CC cross section
errors can be safely taken $\sim 1$ in the SNO analysis.}
Finally, the various SNO background components \cite{HOWT} are
affected by the so-called neutron capture, neutron background, and
low-energy (LE) background correlated systematics, whose
oscillation-independent effects (included in our analysis) are
reported in \cite{HOWT} and not repeated here.

When all such inputs are included, our fit to SNO data only
provides the allowed regions given in the lower right panel of
Fig.~\ref{fig02}. Such regions appear to be somewhat different
from those found in Ref.~\cite{AhDN}, as discussed in Sec.~IV~B.

\begingroup
\squeezetable
\begin{table}[t]
\caption{\label{snospect} \footnotesize \baselineskip=4.mm
Characteristics of the SNO spectrum, divided in 17 day and 17
night bins (first two columns). The experimental rates \cite{AhNC}
are given in the third and fourth columns, and include signal
(CC+ES+NC) and background events. Our evaluation for the
corresponding CC, ES, and NC  components and of their $^8$B $\nu$
shape, energy scale and resolution errors (in the absence of
oscillation, day+night) are given in the remaining 12 columns. The
NC signal is negligible in the last 8 bins. All the quoted numbers
refer to the total $\Phi_B+\Phi_\mathrm{hep}$ contributions. In
the analysis, however, the $\Phi_B$ and $\Phi_\mathrm{hep}$
contributions and their errors are separately evaluated,
analogously to the SK spectrum. In the presence of oscillations,
the last twelve columns are split into day and night contributions
and updated in each point of the parameter space. Backgrounds are
oscillation-independent and are treated according to \cite{HOWT}.
}
\begin{ruledtabular}
\begin{tabular}{ccrr|rrrr|rrrr|rrrr}
Bin & $T$ range & $R^\mathrm{expt}_\mathrm{day}$ &
$R^\mathrm{expt}_\mathrm{night}$ & $R^\mathrm{theor}_\mathrm{CC}$
& \multicolumn{3}{c|}{fractional errors$\times 100$} &
$R^\mathrm{theor}_\mathrm{ES}$ &  \multicolumn{3}{c|}{fractional
errors$\times 100$} &$R^\mathrm{theor}_\mathrm{NC}$ &
\multicolumn{3}{c}{fractional errors$\times 100$} \\
$m$ & (MeV) & (1/kty) & (1/kty) & (1/kty) & shape & scale &resol.
& (1/kty) & shape & scale & resol. & (1/kty) &
shape & scale & resol.\\
\hline
1&$[5,5.5]$
&351.1&399.7&822.2&0.42&$-2.30$&0.10   &135.8&0.64&0.26 &$-0.26$&293.8&1.55&0.88 &$-4.33$\\
2&$[5.5,6]$
&330.9&313.4&875.6&0.63&$-1.89$&$-0.21$&121.7&0.77&0.57 &$-0.35$&248.5&1.55&3.73 &$-2.84$\\
3&$[6,6.5]$
&299.6&272.2&903.0&0.86&$-1.40$&$-0.53$&107.8&0.91&0.93 &$-0.46$&172.1&1.55&7.05 &   0.43\\
4&$[6.5,7]$
&222.4&249.6&902.0&1.12&$-0.85$&$-0.83$& 94.1&1.07&1.32 &$-0.54$& 97.7&1.55&10.85&   5.43\\
5&$[7,7.5]$
&191.2&235.0&872.8&1.41&$-0.22$&$-1.09$& 81.0&1.24&1.77 &$-0.58$& 45.4&1.55&15.16&  12.18\\
6&$[7.5,8]$
&148.9&176.6&817.7&1.73&0.51   &$-1.30$& 68.7&1.44&2.28 &$-0.58$& 17.3&1.55&19.96&  20.68\\
7&$[8,8.5]$
&128.7&122.1&740.9&2.08&1.33   &$-1.42$& 57.2&1.66&2.85 &$-0.54$&  5.4&1.55&25.26&  31.02\\
8&$[8.5,9]$
&139.7&134.1&648.5&2.47&2.25   &$-1.41$& 46.9&1.90&3.48 &$-0.43$&  1.4&1.55&31.19&  43.38\\
9&$[9,9.5]$
& 90.1& 95.6&547.4&2.89&3.29   &$-1.24$& 37.7&2.16&4.19 &$-0.23$&  0.3&1.55&37.69&  58.11\\
10&$[9.5,10]$
& 82.7& 86.3&444.9&3.36&4.45   &$-0.85$& 29.6&2.45&4.99 &0.07   &    0&--- &---  &---    \\
11&$[10,10.5]$
& 66.2& 62.4&347.5&3.86&5.74   &$-0.18$& 22.8&2.77&5.87 &0.51   &    0&--- &---  &---    \\
12&$[10.5,11]$
& 49.6& 59.8&260.4&4.40&7.16   &0.82   & 17.1&3.12&6.86 &1.13   &    0&--- &---  &---    \\
13&$[11,11.5]$
& 31.3& 41.2&186.8&4.97&8.70   &2.21   & 12.5&3.51&7.94 &1.97   &    0&--- &---  &---    \\
14&$[11.5,12]$
& 18.4& 21.2&128.2&5.56&10.37  &4.04   &  8.9&3.92&9.14 &3.07   &    0&--- &---  &---    \\
15&$[12,12.5]$
&  9.2& 18.6& 84.0&6.17&12.15  &6.35   &  6.1&4.36&10.45&4.47   &    0&--- &---  &---    \\
16&$[12.5,13]$
& 11.0& 15.9& 52.6&6.79&14.01  &9.13   &  4.1&4.83&11.87&6.23   &    0&--- &---  &---    \\
17&$[13,20]$
& 9.2 & 9.3 & 71.1&7.68&17.83 &16.34   &  6.5&5.97&15.71&12.48  &    0&--- &---  &---    %
\end{tabular}
\end{ruledtabular}
\end{table}
\endgroup

\section{Quasi-model-independent analysis of SK and SNO}
\label{qmianalysis}

In this last appendix we present a new version of the
model-independent comparison of SK ES and SNO (CC and NC) total
rates, first proposed in \cite{Vill} and then applied in
\cite{Sudb,Adde} to earlier SNO CC data.

For the current SNO threshold $T_\mathrm{SNO} \geq 5$~MeV
\cite{AhNC}, the SK threshold equalizing the SNO CC and SK ES
response functions \cite{Vill} turns out to be $T_{\rm SK}\geq
6.74$ MeV, corresponding to a total electron energy $E_\mathrm{SK}
\geq 7.25$ MeV,
\begin{equation}
\label{equalize} \mathrm{SNO\ CC\ response}\ (T_\mathrm{SNO}\geq
5\mathrm{\ MeV}) \Leftrightarrow \mathrm{SK\ ES\ response}\
(E_\mathrm{SK}\geq 7.25\mathrm{\ MeV})\ .
\end{equation}
In calculating the response functions, we have taken into account
the latest evaluation of the energy resolution width $\sigma_T$ in
SNO \cite{AhNC},
\begin{equation}
\label{sigmasno}
 \sigma_T(\mathrm{SNO}) = -0.0684 + 0.331 \sqrt{T'} +
0.0425\, T'\ ,
\end{equation}
where $T'$ is the true electron kinetic energy, and both $T'$ and
$\sigma_T$ are given in MeV. We have cast, in the same form, the
latest evaluation of $\sigma_T$ for SK (graphically reported in
\cite{Sm01}),
\begin{equation}
\label{sigmask}
 \sigma_T(\mathrm{SK}) = 0.25 + 0.20 \sqrt{T'} +
0.06\, T'\ .
\end{equation}

From the spectral information discussed in App.~\ref{skspectrum}
we estimate (derivation omitted)
\begin{equation}
\label{SKES}
 \Phi_\mathrm{ES}^\mathrm {SK}\simeq 2.35 \pm 0.029 \pm
0.080 \pm 0.045\ (E\geq 7.25\mathrm{\ MeV})\ ,
\end{equation}
in units of $10^6\;\mathrm{cm}^{-2}\,\mathrm{s}^{-1}$. In the
above equation, the first error is statistical, while the second
represents the (properly propagated) sum of systematic errors,
with the exception of the $^8$B shape uncertainty, given
separately as a third error component.

While the above SK ES flux estimate is safe for model-independent
analyses, the current SNO CC flux \cite{AhNC} is not, since its
value {\em does depend\/} on model-dependent assumptions,
governing the CC spectrum distortions and thus the CC/NC
separation \cite{HOWT}. However, we can resort to a
quasi-model-independent analysis, by assuming that the only
observable effect of new neutrino physics can be embedded in a
shift of the first spectral moment \cite{Mome}, namely, in a
linear distortion (tilt) of the CC energy spectrum.%
\footnote{The numerical results for a linear distortion of the CC
component of the SNO spectrum are very similar to those obtained
for a linear distortion of the CC+ES components (not shown), since
the ES contribution is relatively small in SNO.}
This assumption is reasonably general, since the current SNO
statistics is not high enough to really constrain higher moments,
and since we know from SK that only the scenarios with mild
distortions can survive. In this case, the
normalization-preserving form of a generic linear distortion for
the CC differential energy spectrum reads
\begin{equation}
\frac{d N_\mathrm{CC}(T)}{dT}\to \frac{d N_\mathrm{CC}(T)}{dT}
\left( 1 + \frac{T-\langle T\rangle}{\langle T^2\rangle-\langle
T\rangle^2}\, \Delta \langle T\rangle \right)\ ,
\end{equation}
where $\langle T\rangle$ and $\langle T^2\rangle$ are the first
and second moment of the undistorted spectrum, and $\Delta \langle
T \rangle$ is the shift in the first moment.

Assuming  $\Delta \langle T\rangle$ free, our fit to the SNO
energy spectrum gives the following CC and NC fluxes,
\begin{eqnarray}
\Phi^\mathrm{SNO}_\mathrm{CC}
&=& 1.78 \pm 0.10 (\mathrm{stat.})\ ,\label{ccsnoqmi}\\
\Phi^\mathrm{SNO}_\mathrm{NC} &=& 4.90 \pm 0.66(\mathrm{stat.})\
,\label{ncsnoqmi}
\end{eqnarray}
with correlation $\rho=-0.8$. By setting $\Delta\langle
T\rangle=0$, we would instead obtain
$\Phi^\mathrm{SNO}_\mathrm{CC} = 1.73 \pm 0.06 (\mathrm{stat.})$
and $\Phi^\mathrm{SNO}_\mathrm{NC} = 5.29 \pm
0.43(\mathrm{stat.})$ with correlation $\rho=-0.62$, in good
agreement with the results of \cite{AhNC,AhDN} for undistorted
spectrum. The larger statistical errors in Eqs.~(\ref{ccsnoqmi})
and (\ref{ncsnoqmi}) are the price to pay to allow possible linear
distortions in the fit.

Systematic errors are attached as follows. From the (positively
and negatively correlated) SNO error components reported in
Table~II of \cite{AhNC}, we estimate the (CC,NC) experimental
systematics as $(5.2\%,8.8\%)$, with correlation $\rho=-0.5$. The
corresponding theoretical cross section uncertainties
[(1.8\%,1.3\%) from Table~II of \cite{AhNC}] are assumed to have
$\rho\simeq 1$. Finally, if we repeat to the fit leading to
Eqs.~(\ref{ccsnoqmi}) and (\ref{ncsnoqmi}) with a $+1\sigma$
perturbed $^8$B neutrino energy spectrum, we obtain the variations
$(-1.1\%,+3.3\%)$, which are fully correlated among themselves and
with the third SK error component in Eq.~(\ref{SKES}).

In conclusion, we get (in units of
$10^6\;\mathrm{cm}^{-2}\,\mathrm{s}^{-1}$),
\begin{eqnarray}
\Phi_\mathrm{ES}^\mathrm{SK} &=& 2.35 \pm 0.10\ ,\label{phiESSK}\\
\Phi_\mathrm{CC}^\mathrm{SNO} &=& 1.78 \pm 0.14\ ,\label{phiCCSNO}\\
\Phi_\mathrm{NC}^\mathrm{SNO} &=& 4.90 \pm 0.80\ ,\label{phiNCSNO}
\end{eqnarray}
with correlation matrix
\begin{equation}
\rho = \left(\begin{array}{ccc}
  1 & -0.07 & +0.09 \\
   & 1 & -0.65 \\
   &  & 1
\end{array}\right)\ .\label{correl3}
\end{equation}

The above SK and SNO equalized fluxes can be useful to constrain
generic models of new  physics (alternative to---or coexisting
with---usual mass-mixing oscillations), as far as their main
distortion effect on the SNO CC spectrum is approximately linear
in $T$. Within this quasi-model-independent assumption, and in the
hypothesis of active oscillations, the fluxes in
Eqs.~(\ref{phiESSK})--(\ref{phiNCSNO}) are still linked  by the
{\em exact\/} relations \cite{Vill}
\begin{eqnarray}
\Phi_\mathrm{ES}^\mathrm{SK} &=& \Phi_B [\langle P_{ee}\rangle + r
(1-\langle P_{ee}\rangle)] \ ,\label{phiessk}\\
\Phi_\mathrm{CC}^\mathrm{SNO} &=& \Phi_B \langle P_{ee}\rangle\ ,\label{phiccsno}\\
\Phi_\mathrm{NC}^\mathrm{SNO} &=& \Phi_B \ ,\label{phincsno}
\end{eqnarray}
where $P_{ee}(E_\nu)$ is not necessarily constant in energy, and
$r=0.154$ is the ratio of $\nu_{\mu,\tau}$ and $\nu_e$ CC cross
sections (averaged over the current SNO equalized response
function). In the above equations, $\Phi_B$ is the true $^8$B flux
from the Sun (generally different from the SSM value), and
$\langle P_{ee}\rangle$ is the energy average of $P_{ee}(E_\nu)$
over the response function (equal in SK and SNO).

\begin{figure}
\includegraphics[scale=0.88, bb= 100 100 500 720]{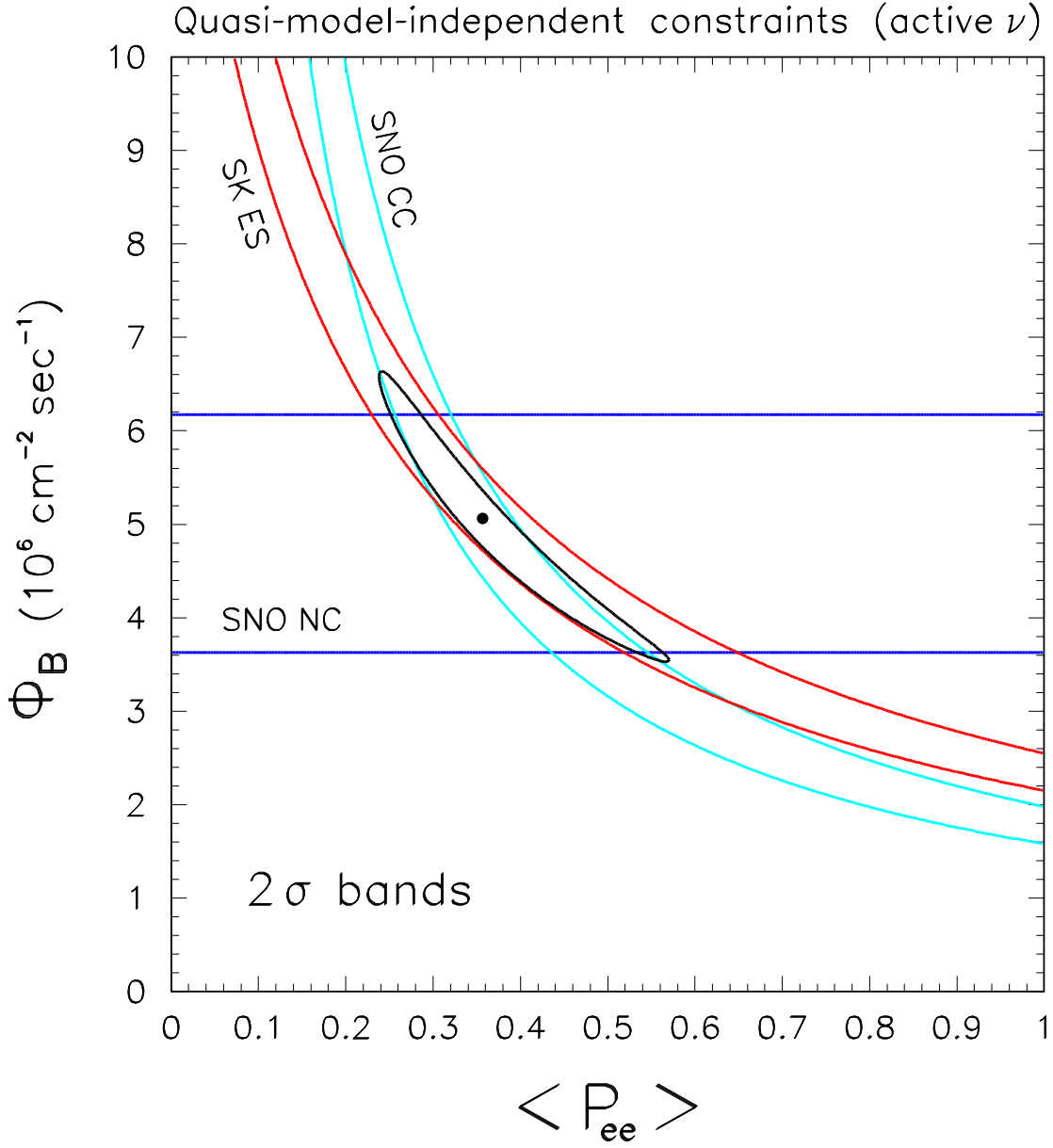}
\vspace*{+0cm} \caption{ \label{fig09} Quasi-model-independent
comparison of SK (ES) and SNO (CC and NC) total event rates in the
plane charted by the boron flux $\Phi_B$ and by the average
$\nu_e$ survival probability $\langle P_{ee}\rangle$, for
equalized SK and SNO response functions. The evaluation of the SNO
rates includes possible {\em linear\/} distortions in the energy
spectrum. The $2\sigma$ bands for each datum $(\Delta \chi^2=4)$
appear to be in very good agreement with each other for
$P_{ee}\sim 1/3$ and for $\Phi_B$ close to its SSM prediction
\cite{Ba01}. The combination of the $2\sigma$ bands is also shown
(slanted elliptical region).}
\end{figure}

Equations~(\ref{phiESSK})--(\ref{correl3}) and
(\ref{phiessk})--(\ref{phincsno}) overconstrain the two parameters
$\Phi_B$ and $\langle P_{ee}\rangle$. Figure~\ref{fig09} shows, in
($\Phi_B,\langle P_{ee}\rangle$) coordinates, both separate and
combined bounds at the $2\sigma$ level for each datum ($\Delta
\chi^2=4$). There is very good agreement between any two out of
the three data in Eqs.~(\ref{phiESSK})--(\ref{phiNCSNO}). Their
combination strengthens previous model-independent indications
\cite{Sudb,Adde} for a consistency of $\Phi_B$ with the SSM
prediction \cite{Ba01} and for $\langle P_{ee}\rangle \sim 1/3$ in
the SK-SNO energy range.





\end{document}